\documentclass[preprint]{aastex}
\usepackage{subfigure}
\usepackage{graphicx}
\usepackage{graphics}
\usepackage{natbib}
\newcommand{\wdotactual}{$\dot\omega_{\rm tot} = 0.07089^{+0.00021}_{-0.00013}$ deg~cycle$^{-1}$}

\newcommand{\eccvalue}{ $0.07755^{+0.00018}_{-0.00026}$} 
\newcommand{\wdotvalue}{$0.07089^{+0.00021}_{-0.00013}$}
\newcommand{\uvalue}{$33.48^{+0.10}_{-0.06}$}

\newcommand{\mrGenevaDiff}{1.6} 
\newcommand{\mrUtrechtDiff}{0.4}
\newcommand{\mrGranadaDiff}{0.2} 

\newcommand{\gtGenevaDiff}{0.1} 
\newcommand{\gtUtrechtDiff}{3.5\pm1.5}
\newcommand{\gtGranadaDiff}{5.5\pm1.0}

\newcommand{\bestIncl}{$     72.09\pm      0.06$}
\newcommand{\bestOone}{$      4.88\pm      0.03$}
\newcommand{\bestOtwo}{$      4.89\pm      0.04$}
\newcommand{\bestRone}{$      5.41\pm      0.04$}
\newcommand{\bestRtwo}{$      4.29\pm      0.05$}
\newcommand{\bestTtwo}{$    25750\pm      435$}
\newcommand{\bestTone}{$    30000\pm      500$}
\newcommand{\bestloggone}{$     4.133\pm     0.018$}
\newcommand{\bestloggtwo}{$     4.185\pm     0.021$}
\newcommand{\bestTrat}{$     0.858\pm     0.002$}
\newcommand{\bestFone}{$      1.08\pm      0.04$}
\newcommand{\bestFtwo}{$      1.10\pm      0.03$}
\newcommand{\bestMone}{$     14.54\pm      0.08$}
\newcommand{\bestMtwo}{$     10.29\pm      0.06$}
\newcommand{\bestvone}{$      123\pm        5$}
\newcommand{\bestvtwo}{$       99\pm        3$}
\newcommand{\bestLone}{$      4.33\pm      0.03$}
\newcommand{\bestLtwo}{$      3.86\pm      0.03$}
\newcommand{\bestchisqr}{$   3489.00$}

\newcommand{\TfracSp}{$0.0012$}
\newcommand{\InclSp}{$0.2$}
\newcommand{\OmegaOneSp}{$0.045$}
\newcommand{\OmegaTwoSp}{$0.045$}

\newcommand{\OmegaOneSpF}{$0.005$}
\newcommand{\OmegaTwoSpF}{$0.005$}
\newcommand{\TfracSpF}{$0.03$}
\newcommand{\InclSpF}{$0.0005$}

\newcommand{\bestlogktwonewt}{$-1.975\pm0.017$}
\newcommand{\bestlogktwogr}{$-3.412\pm0.018$}
\newcommand{\bestwdotnewt} {$0.06830\pm0.00017$}
\newcommand{\bestwdotgr} {$0.002589\pm0.000015$}

\newcommand{\theowdotnewt}{$0.06883\pm0.00017$}
\newcommand{\theologktwonewt}{$-2.005\pm0.025$}

\newcommand{\Stromulf}{$     0.710\pm     0.007$}                 
\newcommand{\Stromvlf}{$     0.690\pm     0.008$}                 
\newcommand{\Stromblf}{$     0.685\pm     0.007$}                 
\newcommand{\Stromylf}{$     0.683\pm     0.007$}                 
\newcommand{\JohnsUlf}{$     0.706\pm     0.008$}                 
\newcommand{\JohnsBlf}{$     0.689\pm     0.007$}                 
\newcommand{\JohnsVlf}{$     0.683\pm     0.007$}

\newcommand{\Hermeslf}{$0.700\pm0.02$}

\newcommand{\cosIncl}{$72.15$}

\newcommand{\cosOone}{$4.92$}
\newcommand{\cosOtwo}{$4.89$}

\newcommand{\cosTrat}{$0.860$}
\newcommand{\coschisqr}{$3480.01$}

\newcommand{\logIncl}{$72.14$}

\newcommand{\logOone}{$4.91$}
\newcommand{\logOtwo}{$4.88$}

\newcommand{\logTrat}{$0.858$}
\newcommand{\logchisqr}{$3503.96$}

\newcommand{\ldIncl}{$72.18$}

\newcommand{\ldOone}{$4.92$}
\newcommand{\ldOtwo}{$4.89$}

\newcommand{\ldTtwo}{$25049$}
\newcommand{\ldTrat}{$0.835$}
\newcommand{\ldchisqr}{$3299.11$}

\newcommand{\ThirdIncl}{$72.24$}

\newcommand{\ThirdOone}{$4.94$}
\newcommand{\ThirdOtwo}{$4.87$}

\newcommand{\ThirdTrat}{$0.855$}
\newcommand{\Thirdchisqr}{$3414.93$}

\newcommand{\refIncl}{$72.20$}

\newcommand{\refOone}{$4.90$}
\newcommand{\refOtwo}{$4.92$}

\newcommand{\refTrat}{$0.856$}
\newcommand{\refchisqr}{$3522.57$}

\newcommand{\ThiIncl}{$72.15$}

\newcommand{\ThiOone}{$4.92$}
\newcommand{\ThiOtwo}{$4.89$}

\newcommand{\ThiTrat}{$0.867$}
\newcommand{\Thichisqr}{$3488.01$}

\newcommand{\TLowIncl}{$72.17$}
\newcommand{\TLowOone}{$4.94$}
\newcommand{\TLowOtwo}{$4.87$}
\newcommand{\TLowTrat}{$0.857$}
\newcommand{\TLowchisqr}{$3460.16$}

\newcommand{\HensIncl}{$72.58\pm0.30$}
\newcommand{\HensRone}{$5.23\pm0.06$}
\newcommand{\HensRtwo}{$4.32\pm0.07$}
 
\newcommand{\HensTtwo}{$26400\pm400$}
\newcommand{\HensTrat}{$0.88\pm0.020$}
\newcommand{\Hensecc}{$0.0867\pm0.0006$} 
\newcommand{\Hensomega}{$153.3\pm0.6$}
\newcommand{\Hensvonesini}{$117\pm4$}
\newcommand{\Hensvtwosini}{$94\pm2$}

\def\astrosun {\mbox{$\odot$}}
\newcommand{\Msol}{\ensuremath{\mbox{M}_{\astrosun}}}
\newcommand{\Rsol}{\ensuremath{\mbox{R}_{\astrosun}}}
\newcommand{\Lsol}{L_{\odot}}

\usepackage{natbib}
\usepackage{graphicx,epsfig}
\usepackage{bm}
\usepackage{color}
\usepackage{comment} 

\shorttitle{Absolute dimensions of V578~Mon} 
\shortauthors{Garcia et al.}

\begin{document}
\title{A Strict Test of Stellar Evolution Models: The Absolute Dimensions of Massive Benchmark Eclipsing Binary V578~Mon}
\author{E.V.\ Garcia\altaffilmark{1,2}, Keivan G.\ Stassun\altaffilmark{1,3}, 
K.\ Pavlovski\altaffilmark{4},H.\ Hensberge\altaffilmark{5},
Y.\ G\'omez Maqueo Chew\altaffilmark{6}, 
A.\ Claret\altaffilmark{7}}
\altaffiltext{1}{Department of Physics \& Astronomy, Vanderbilt University, 
6301 Stevenson Center, VU Station B 1807, Nashville, TN 37235; eugenio.v.garcia@gmail.com}
\altaffiltext{2}{Lowell Observatory Pre-doctoral Fellow, Lowell Observatory, Flagstaff, AZ 86001}
\altaffiltext{3}{Department of Physics, Fisk University,
1000 17th Ave.\ N., Nashville, TN, USA 37208}
\altaffiltext{4}{Department of Physics, University of Zagreb, Bijeniÿcka cesta 32, 10000 Zagreb, Croatia}
\altaffiltext{5}{Royal Observatory of Belgium, Ringlaan 3, 1180 Brussels, Belgium}
\altaffiltext{6}{Physics Department, University of Warwick, Gibbet Hill Road, Coventry CV4 7AL, United Kingdom}


\altaffiltext{7}{Instituto de Astrof'sica de Andaluc'a, CSIC, Apartado 3004, 18080, Granada, Spain}

\begin{abstract} 
We determine the absolute dimensions of the eclipsing binary V578~Mon, 
a detached system of two early B-type stars (B0V + B1V, P$=$2.40848 d) in the 
star-forming region NGC~2244 of the Rosette Nebula. From the light curve 
analysis of 40 yr of photometry and the analysis of {\sc hermes} spectra, 
we find radii of $5.41\pm0.04$~Rsun~and $      4.29\pm      0.05$~Rsun, 
and temperatures of $    30000\pm      500$~K and $    25750\pm      435$~K respectively. 
We find that our disentangled component spectra for V578~Mon agree 
well previous spectral disentangling from the literature. 
We also reconfirm the previous spectroscopic orbit 
of V578~Mon finding that masses of $     14.54\pm      0.08$~Msun~and 
$     10.29\pm      0.06$~Msun~are fully compatible with the new analysis.  
We compare the absolute dimensions to the rotating models of the 
Geneva and Utrecht groups and the models of Granada group. We find all three 
sets of models marginally reproduce the 
absolute dimensions of both stars with a common age 
within uncertainty for gravity-effective temperature isochrones. 
However - there are some apparent 
age discrepancies for the corresponding mass-radius isochrones. Models with larger 
convective overshoot $>0.35$ worked best. 
Combined with our previously determined apsidal motion of 
$0.07089^{+0.00021}_{-0.00013}$ deg~cycle$^{-1}$, we compute the internal structure 
constants (tidal Love number) for the newtonian and general relativistic 
contribution to the apsidal motion, $\log{k_2}=-1.975\pm0.017$
and $\log{k_2}=-3.412\pm0.018$ respectively. We find the relativistic 
contribution to the apsidal motion of be small $<4\%$. 
We find that the prediction of $\log{k_{\rm 2,theo}}=-2.005\pm0.025$
of the Granada models fully agrees with our observed $\log{k_2}$. 
\end{abstract}


\section{Introduction\label{sec:intro}}

\begin{figure}[ht]
\begin{center}
\includegraphics[angle=90,width=\textwidth]{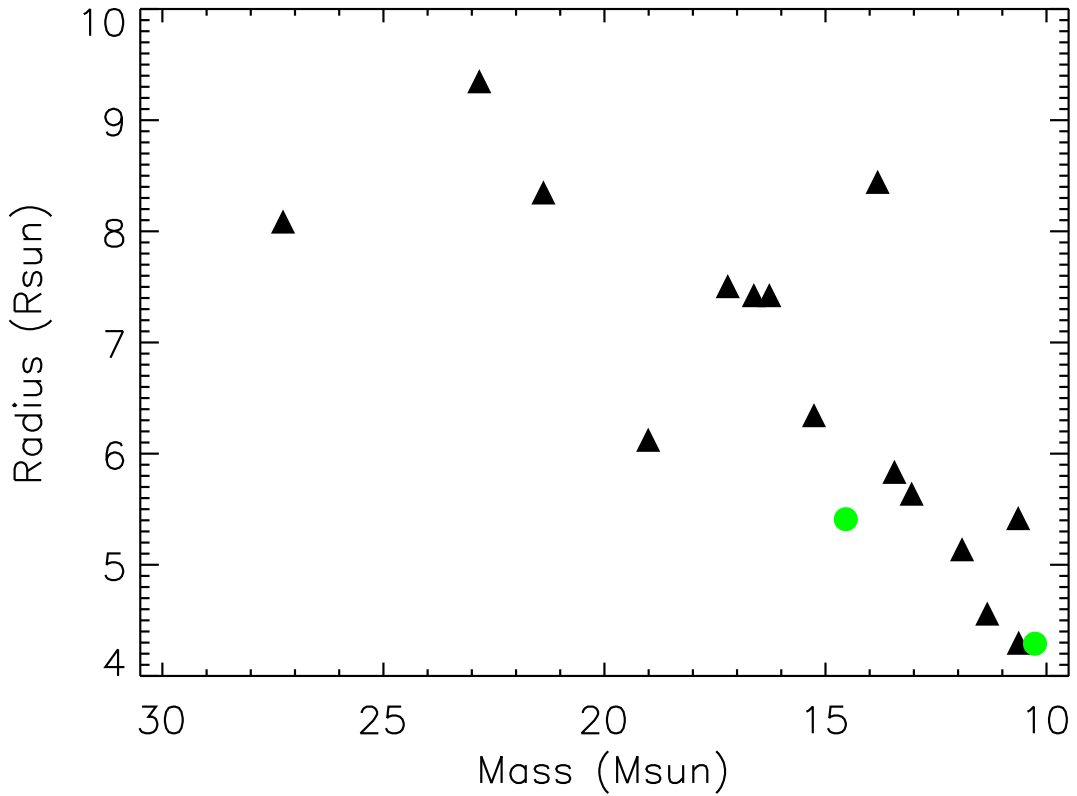}
\end{center}
\caption{\label{fig:hr} Massive ($>10\Msol$) detached eclipsing binaries 
with accurate masses and radii better than $2\%$ are scarce. There are only 
9 such systems (black triangles) including V578~Mon (green circles). 
This list of eclipsing binaries is adapted from \cite{Torres10}. The error bars 
on the mass and radii are smaller 
than the plotted symbols. Of these eclipsing binaries, 
V578~Mon is simultaneously one of the 
youngest and lowest mass ratios $q=\frac{M_2}{M_1}$. 
}
\end{figure}

Detached eclipsing binary stars (dEBs) provide accurate 
observed stellar masses, radii, 
effective temperatures, and rotational velocities.  
See a recent review by \cite{Torres10} 
for a discussion of 94 dEBs with accurate masses and radii used to 
test stellar evolution models. There are only nine 
total massive dEBs,  or equivalently 18 stars 
whose physical parameters have been 
determined with an accuracy of better than 3\%, making 
V578~Mon one of only nine EBs with $M_1 \ge M_2 > 10$ \Msol\ and with 
sufficient accuracy to be included in the 
\citet{Torres10} compilation of benchmark-grade EBs. 
Figure~\ref{fig:hr} demonstrates the upper main 
sequence of all dEBs with $M_1 \ge M_2 > 10$ \Msol\ 
and masses and radii determined to 3\% (adapted from \cite{Torres10}). 
V578~Mon is therefore a benchmark system for testing stellar 
evolution models of newly formed massive stars. The accurate absolute dimensions of 
eclipsing binary stars provide a unique opportunity to 
test stellar evolution models in two 
ways: the ``isochrone test" and the``apsidal motion test''.

The ``isochrone test'' of stellar evolution models requires 
that the ages of both components of a dEB predicted 
from separate stellar evolution tracks to be 
the same within uncertainty of the 
absolute dimensions $(M, R, T_{\rm eff}, v_{\rm rot})$. 
For the ``isochrone test'', we assume that both components 
of the dEB formed together in the same initial gas cloud. Therefore,
both components of a dEB are assumed to arrive at the
zero age main sequence (ZAMS) at nearly the same time. 
Furthermore,
their initial chemical compositions must be the same. Finally, we assume
that each component of the binary 
evolves in isolation, where the 
effects of the companion star on the evolution is small 
or negligible.

The ``isochrone test'' is strongest for eclipsing binaries 
with low mass ratios $q=\frac{M_2}{M_1} < 1$. 
For dEBs where component masses $M_1\approx M_2$, both 
stars will evolve on the same evolutionary track. This 
does not allow for strict tests of stellar evolution models unless 
the chemical composition or effective temperature of the stars 
is known.  Stellar evolution models 
will predict two stars of the same mass and composition to 
have the same age. 
Conversely, the larger the difference in 
initial mass between the components of the binary star, 
the larger the difference in main sequence (MS) lifetimes of the two stars. 
Therefore, the stellar models must have the accurate input physics
to correctly predict how quickly stars of different mass evolve relative 
to each other. The correct input physics in turn 
yields correct predictions of the 
observed absolute dimensions of the 
detached eclipsing binary. 

Detached eclipsing binary stars with apsidal motion (precession of the argument of periastron)
also allow for the ``apsidal motion test'' of the stellar internal structure \citep{Claret10}. 
Physically, the observed apsidal motion rate in an eclipsing binary is a result 
of the tidal forces of each star on each other. In turn, this tidal force is linked to 
the internal structure of each star, the star's separation, their mass ratio $q$ and 
their radii $R_1$ and $R_2$. The internal 
structure is quantified by the constant $\log{k_2}$ which is the logarithm of 
 twice the tidal Love number \citep{Kramm11}. 
The ``apsidal motion'' test compares the theoretical internal 
structure constant $\log{k_{\rm 2,theo}}$ to the observed internal 
structure constant $\log{k_{\rm 2,obs}}$. The observed internal structure
constant is a function the observed absolute dimensions and 
apsidal motion of the eclipsing binary. The observed internal structure 
constant is very sensitive to the radii ($k_{\rm 2,obs}\propto R^{5}$) - therefore,
this test can only be performed with accurate stellar radii. However, including this 
study of massive dEB V578~Mon, there are only five massive, eccentric eclipsing 
binaries available for these tests of internal structure \citep{Claret10}. 

Here we combine the previous 
determination of $\dot\omega$ and $e$ from \cite{Garcia11} with a 
reanalysis of 40 years worth of photometry to 
re-determine the fundamental properties of V578~Mon. We include the photometry 
used by the previous light curve analysis \citep{H2000}. 
We compare the masses, temperatures and radii of V578~Mon with rotating high mass stellar 
evolution models by Granada \citep{Claret04,Claret06}, Geneva \citep{Georgy13,Ekstrom12}, 
 and Utrecht \citep{Brott11} groups. We also compare the observed internal structure 
constant $\log{k_{\rm 2,obs}}$ with theoretical 
$\log{k_{\rm 2,theo}}$ using the methods of \cite{Claret10}.

\section{The Eclipsing Binary V578~Mon in NGC 2244 \label{sec:EBV578Mon}}

The photometric variability of the bright (V=8.5), 2.408 day period, 
eccentric, massive detached eclipsing binary
(dEB) V578~Mon (HDE~259135, BD$+4^\circ$1299), comprising a B1V type primary
star and a B2V type secondary
star was first identified in the study by \citet{Heiser77} 
of NGC~2244 within the Rosette Nebula (NGC~2237, NGC~2246). 
The identifications, locations and photometric parameters for V578 Mon 
are listed in Table~\ref{Table:obs}.
The absolute dimensions of V578~Mon have been determined from three seasons
of Str\"{o}mgren $uvby$ photometry and one season of radial-velocity
data by \citet{H2000}. An analysis of
the metallicity and evolutionary status of V578~Mon was undertaken by
 \citet{PavHens05} and \cite{H2000}.
The masses and radii of V578~Mon determined from these data are
\bestMone~\Msol\ and \bestMtwo~\Msol, and 
\HensRone~\Rsol\ and
\HensRtwo~\Rsol\, for the primary and secondary respectively \citep{H2000}.
V578~Mon was included in the list of 94 detached eclipsing binaries 
with masses and radii accurate to $2\%$ by \cite{Torres10}. The 
radii for V578~Mon listed in \cite{Torres10} were found to be incorrect 
by \cite{Garcia13} given system's eccentric orbit and asynchronous rotation. 
The apsidal motion $\dot\omega$ and a new eccentricity $e$ were determined in 
\cite{Garcia11}. V578~Mon was observed by MOST \citep{Pribulla10}.

Given the inclination of V578~Mon, its eclipses are partial, meaning that 
neither star is fully out of view of Earth. Partial eclipses can 
translate into a degeneracy between the radii, preventing the component 
radii $R_1$ and $R_2$ from being individually measured. However, V578~Mon 
also has an eccentric orbit, meaning that the eclipse durations are not equal, 
which helps breaks this degeneracy and allows the radii to be determined separately. 
V578~Mon is observed to not have tidally locked yet. The system has 
a low mass ratio $q$=$0.7078$ as compared to similar 
systems with well-determined absolute parameters such as V1034 Sco, V478 Cyg, AH Cep, 
V453 Cyg, and CW Cep \citep{Bouzid05, PopperEtzel81,PopperHill91,Bell86,
Holmgren90, Southworth04,Popper74,Stickland92}. 
Of all of these systems, V578 Mon is also the youngest, making 
this system a benchmark case for testing stellar evolution models 
at the youngest ages.



\section{Data\label{sec:data}} 

\subsection{Johnson UBV and Str\"{o}mgren $uvby$ Photometry}
The available time-series photometry of V578~Mon covers nearly 40~yr and 
more than one full apsidal motion period. 
A summary of the various light curve epochs, including filters and observing
facilities used, is presented in Table~\ref{Table:Phot}. 
Photometry from \citet{Heiser10}
includes multi-band light curves spanning 1967--2006 from the
16-in telescope at Kitt Peak National Observatory (KPNO) and from the
Tennessee State University(TSU) -Vanderbilt 16-in Automatic 
Photoelectric Telescope (APT) at Fairborn Observatory.  The KPNO
Johnson $UBV$ light curves comprise 725 data points spanning 1967--1984 
with average uncertainties per data point of 0.004 mag computed by \cite{Heiser10}. 
The APT Johnson $BV$ light curves span 1994--2006 and consist
of 1783 data points with average uncertainties per data point 
of 0.001 mag for B and 0.002 mag for V \citep{Heiser10}. Light curves 
from \cite{H2000} span 1991--1994 from the 0.5-m Str\"{o}mgren Automatic Telescope (SAT) 
at La Silla, with 248 data points in each of the $uvby$ filters and average 
uncertainty per data point of 0.003 mag \citep{H2000}. 
We begin our light curve analysis with the observational errors originally 
estimated by \cite{Heiser10} and \cite{H2000}. 
Table~\ref{Table:Phot} lists these average uncertainties,
$\sigma_0$, as reported by the original authors. 
However, from our light curve fits (see below) we found that these 
uncertainties were in most cases underestimated. Thus
we also report as $\sigma$ in Table~\ref{Table:Phot} the 
uncertainties that we ultimately adopted for each light curve.

\subsection{ {\sc hermes} Spectroscopy} 

A new series of high-resolution echelle spectra were secured in
December 2011 (36 exposures) and February 2012 (8 exposures) with
{\sc hermes}, the fiber-fed high-resolution spectrograph on the
Mercator telescope located at the Observatorio del Roque de los
Muchachos, La Palma, Canary Islands. {\sc hermes} samples the entire
optical wavelength range (3800-9000 {\AA}) with a resolution of $R = 85\,000$
\citep{Raskin11}. The observations listed in Table~\ref{Table:Hermes} 
cover the orbital cycle uniformly. 
Groups of two concatenated exposures allow us to obtain
a robust estimation of random noise as a function of wavelength, and a
check on cosmic ray events surviving the detection algorithm in the data
reduction. In total, 44 exposures were obtained at 19 epochs, 16 of which
are out of eclipse. One series of six exposures starts near the primary
mid-eclipse; One series of two concatenated exposures taken around
secondary mid-eclipse has a significantly lower exposure level, but
another one consisting of four concatenated exposures starting around
secondary mid-eclipse is available. 

Exposure times close to 2100~s were used for most spectra, but in case of 
one out-of-eclipse epoch the exposure time was significantly shorter, 
1200~s. The signal-to-noise ratio of the spectra is 50 to 100 at 4000~{\AA}, then 
rapidly increasing to 120 up to 200 at 5000~{\AA} and remaining close to 
this level at longer wavelengths. The numbers apply to the sum of two 
concatenated exposures. The reduction of the spectra has
been performed using the {\sc heres} pipe-line software package. The
spectra resampled directly in constant-size velocity bins (ln $\lambda$),
very nearly in size to the detector pixels, were used. Normalization to
the continuum is done separately.

The {\sc hermes} spectra outnumber the {\sc caspec} spectra used by 
Hensberge et al.\~(2000), but fall short with regard to signal-to-noise 
ratio. However, they cover a much larger wavelength region, include 
epochs in both eclipses and cover the orbit more homogeneously. In the 
wavelength region covered by both sets, the reconstruction has better  
signal-to-noise ratio in the {\sc caspec} set, but the risk of bias due  
to phase gaps might be higher with the {\sc caspec} data. Both data sets 
were obtained in different parts of the apsidal motion cycle.



\section{Analysis \label{sec:lcanly}}


\subsection{Spectral Disentangling \& Light Ratio} 

In V578~Mon binary system the eclipses are partial which causes degeneracy
in the light curve solution for the radii of the components.  It was
checked whether a spectroscopic light ratio has sufficient precision to
reduce the degeneracy. This light ratio might be constrained either by
the changing line dilution during the eclipse, or/and by constrained
fitting of the reconstructed component spectra by theoretical spectra,
simultaneously deriving the light ratio as well as the photospheric parameters
\citep{Tamajo11}. In the latter implementation, the
light ratio is assumed identical in all observed spectra, hence eclipse
spectra are not used.

With partial eclipses of roughly 0.1 mag
depth, and less for the secondary eclipse at the epoch of the spectroscopy,  
line depth in the composite
spectrum is affected at the level of 0.5\% of the continuum only when the
two components have in their intrinsic spectra a line differing by 7\% of
the continuum depth. The similarity of the components and the rotational
broadening in the spectra imply that no metal line approaches this level.
Hence, using the changing line dilution to measure the light ratio 
precisely is challenging. Exceedingly large signal-to-noise ratios 
would be required to be able to use single or few lines. Including 
many lines, i.e. large stretches of spectrum offers the opportunity 
to reduce the requirements on the signal-to-noise ratio. However, 
bias in tracing the continuum is expected to put an upper limit to 
the precision with which the light ratio can be measured in a system 
with components with similar spectra and substantial rotation.


Therefore, we explored the alternative option of constrained fitting,
although it is model-sensitive. Spectral disentangling \citep{Had95}, further 
referred as {\sc spd} is performed in a spectral
range of about 100 - 150 {\AA} (of the order of 4000 bins) in the
wavelength range 3900 - 5000 {\AA},  centered on prominent lines of
He~{\sc i}, He~{\sc ii} and stronger metal lines. The apsidal motion
study  \citep{Garcia11} permitted us to fix the eccentricity $e$,
the longitude of the periastron, $\omega$, for the epoch of the spectra,
and the time of periastron passage. The {\sc spd} code used is
{\sc FDBinary}\footnote{{\sf http://sail.zpf.fer.hr/fdbinary/}} \citep{fdb}. 

\begin{figure}[ht]
\begin{center}
\includegraphics[width=\textwidth]{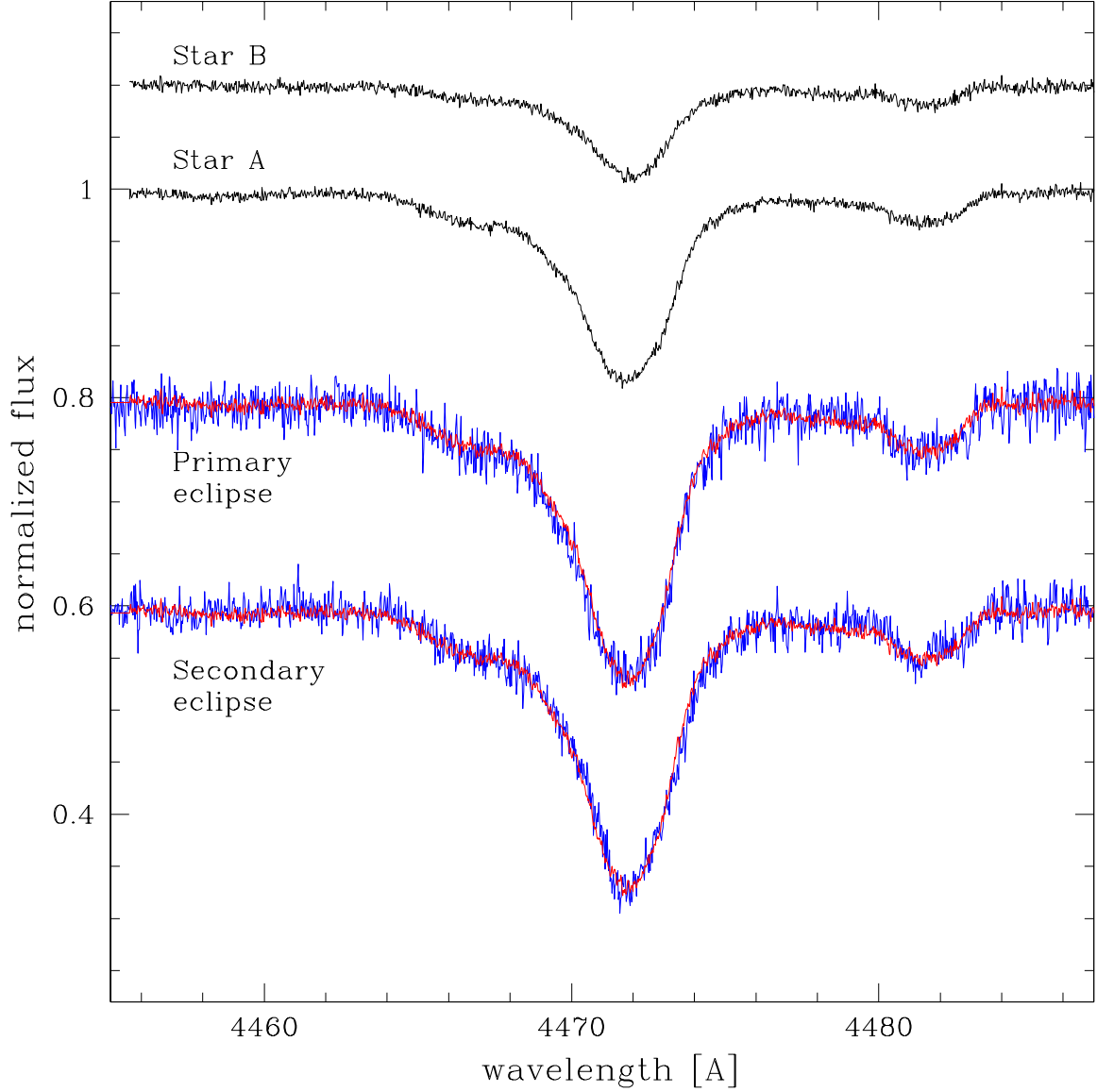}
\end{center}
\caption{\label{fig:eclip} Fits (red) to the {\sc hermes} spectra (blue) obtained during the 
primary and secondary eclipse of V578~Mon. The disentangled component 
spectra obtained from time-series of observed spectra out-of-eclipse are shown 
above in black. The light ratio from the light curve analysis agrees to 
within uncertainty the light ratio derived from the in eclipse spectra. The light 
contribution of each component in the phases of the eclipses were calculated 
from the final light curve solution. 
}
\end{figure}
\begin{figure}[ht]
\centering
\subfigure[He 4388]{%
\includegraphics[width=0.45\linewidth]{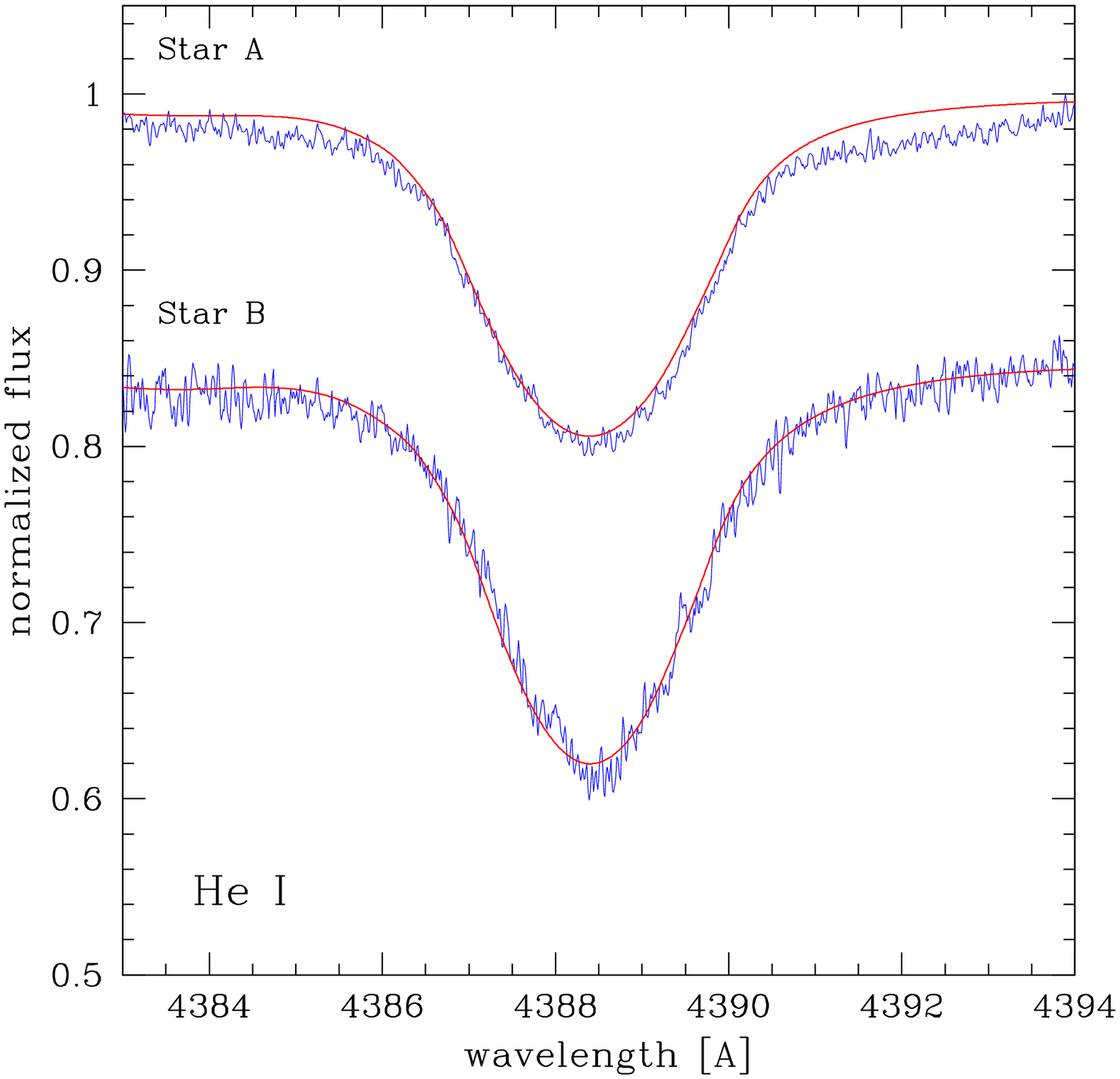}
\label{fig:He1}}
\quad
\subfigure[He 4471]{%
\includegraphics[width=0.45\linewidth]{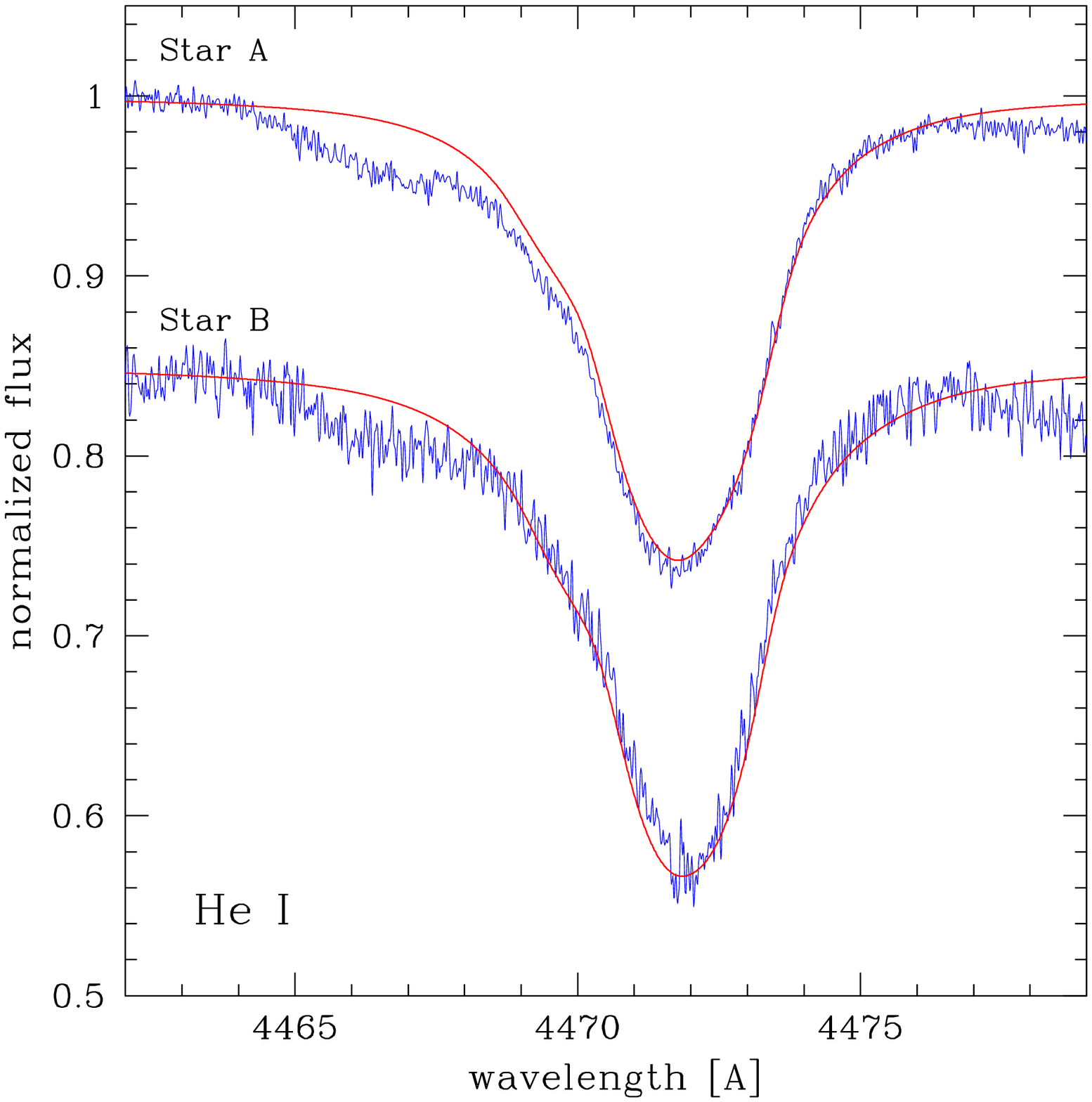}
\label{fig:He2}}
\subfigure[He 4541]{%
\includegraphics[width=0.45\linewidth]{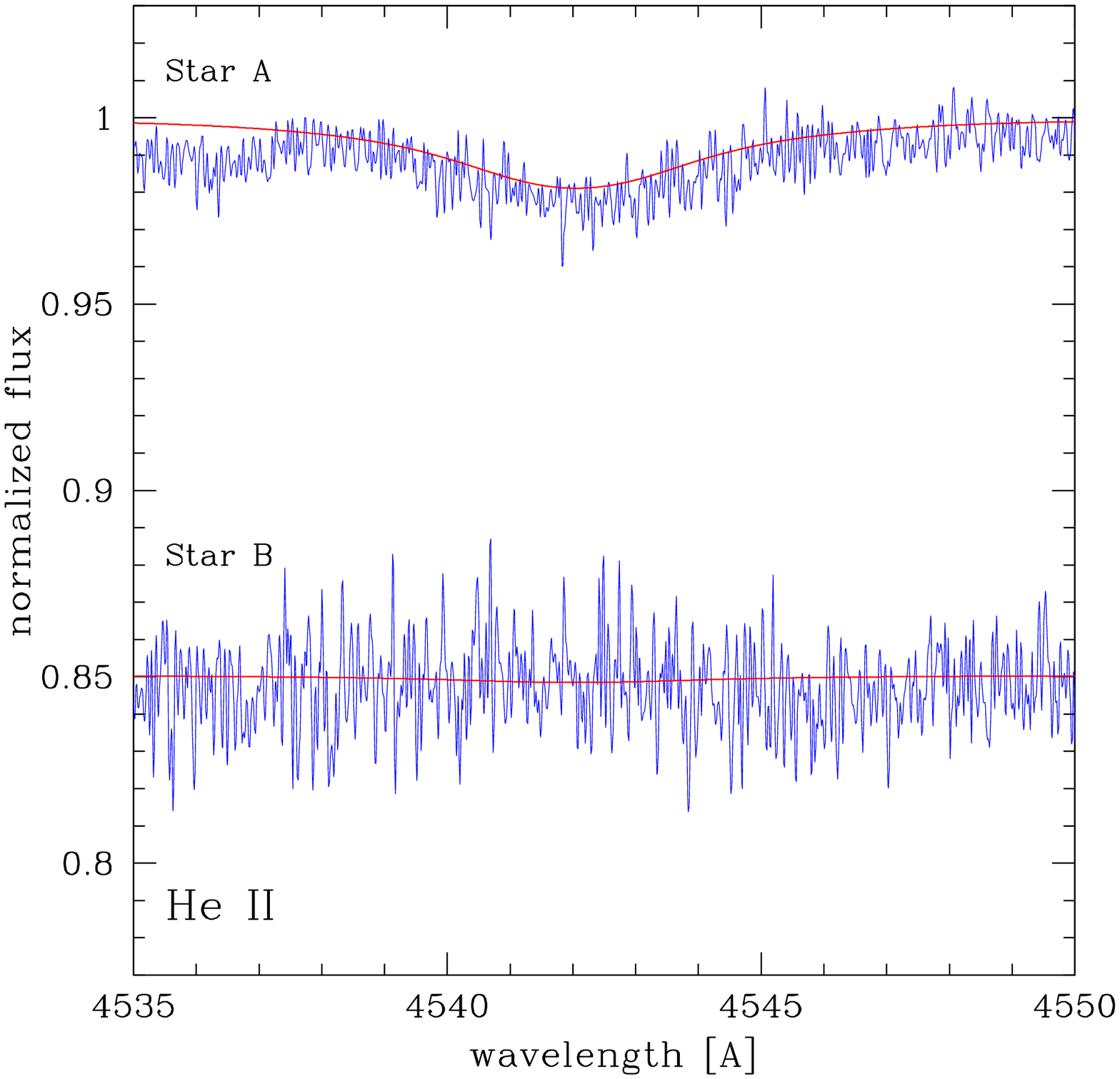}
\label{fig:He3}}
\quad
\subfigure[He 4686]{%
\includegraphics[width=0.45\linewidth]{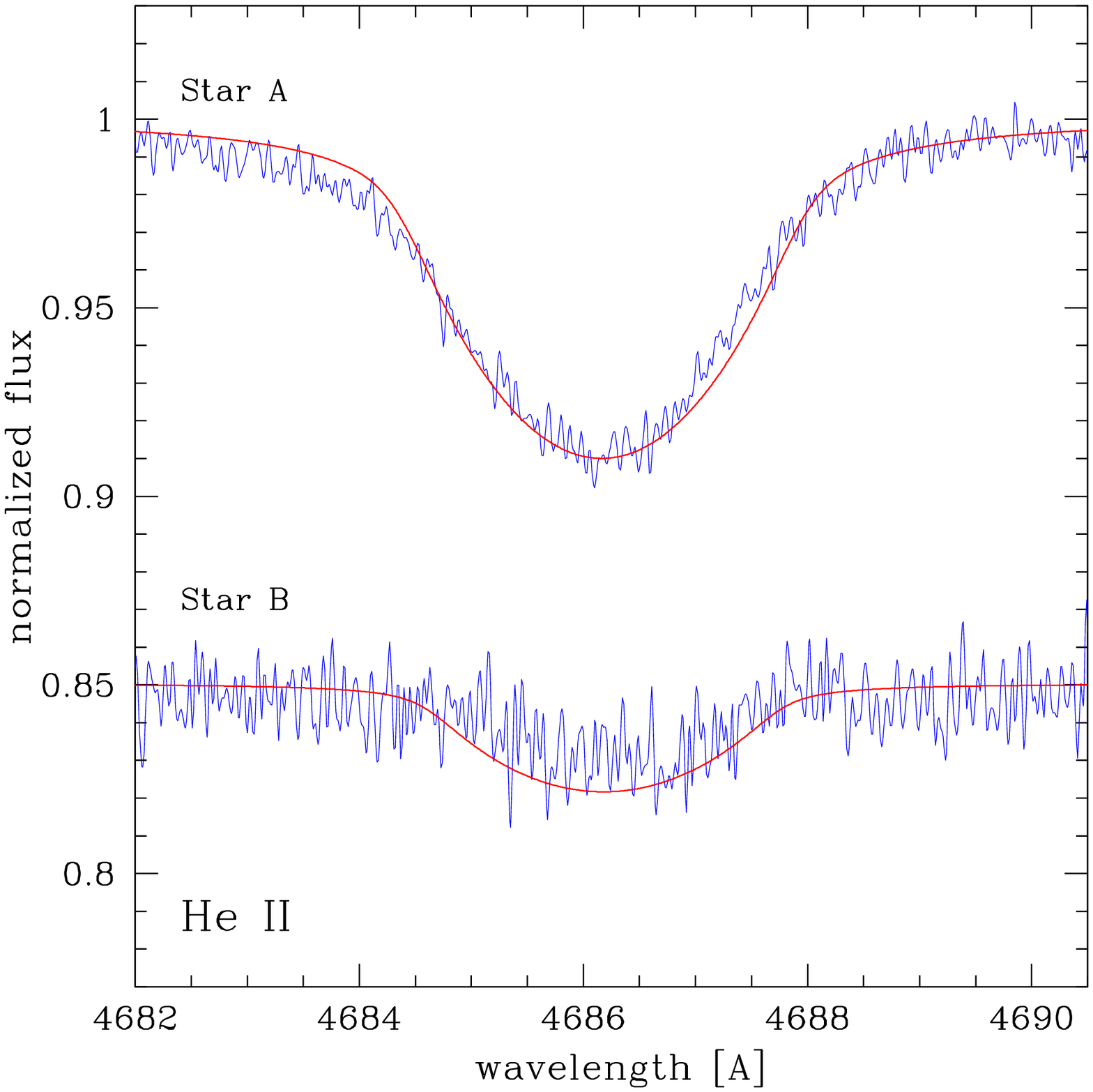}
\label{fig:He4}}
\caption{\label{fig:he} Optimal fitting for the four helium
 lines, He I 4388 {\AA}, He I 4471 {\AA}, He II 4541 {\AA}, and He II 4686 {\AA} for 
 the out-of-eclipse {\sc HERMES} spectra. 
 In each panel helium line profiles for both components are
 shown (blue solid line). Optimal fitting was performed on all 4 lines simultaneously (red solid line). 
 These are reconstructed helium profiles from disentangled
 spectra using the light ratio and surface gravities fixed to the
 final solution. A color version of this figure is available in the online journal. }
\label{fig:figure}
\end{figure}

{\sc spd} applied to selected spectral regions of the {\sc hermes} 
spectra, well distributed over the full range of Doppler shifts in 
the orbit (see orbital phases in Table~\ref{Table:Hermes}), leads to radial velocity 
amplitudes K1 and K2 compatible with \cite{H2000} within  
better than 1 km/s. Thus the spectra are 
reconstructed using the mean orbital elements (Table \ref{Table:RV}), now also
including regions around $H_{\gamma}$ and H$_{\delta}$ ($H_{\beta}$
has a broad interstellar band centered on its red wing). For the
constrained fitting, optimization was done for hydrogen and helium
lines only, and for combinations of them. The reconstructed spectra 
for both out-of-eclipse and in-eclipse phases 
are shown in Figures~\ref{fig:eclip} 
and \ref{fig:he}. 

The component spectra for different dilution factors can be obtained
from a single disentangling computation, followed by an adequate
renormalisation. As starting point for the photospheric parameters,
$T_{\rm eff, 1} = 30\,000$ K, and $T_{\rm eff, 2} = 26\,400$ K is used,
based on the extensive study of \cite{H2000} and \cite{PavHens05}. 
The surface gravities of the components are fixed
to $\log{g_{\rm 1}}$=~\bestloggone~and $\log {g_{\rm 2}}$=~\bestloggtwo~as derived in
this paper. This suppresses the degeneracy of line profiles of hot stars
in the (temperature, gravity) plane.
Calculations for a small grid in $\log{g}$ has shown that the effect
of fixing $\log{g}$ might produce deviations of about a few tenths of
the percentage in determining the light dilution factors.

Optimization of relative light factors includes a search
through a grid of theoretical spectra, using a genetic algorithm. 
A grid of synthetic spectra was calculated assuming non-LTE line
formation. The calculations are based on the so-called
hybrid approach of \cite{Nieva07} in which model atmospheres are
calculated in LTE approximation and non-LTE spectral synthesis with
detailed statistical balance. Model atmospheres are constructed
with {\sc atlas9} for solar metallicity, [M/H] = 0, and helium
abundance by number density, $N_{\rm He}/N_{\rm tot} = 0.089$ \citep{Castelli97}.
Non-LTE level populations and model spectra were computed with
recent versions of  {\sc detail} and {\sc surface} \citep{Butler1985}.
Further details on the method, grid and calculations
 can be found in \cite{Tamajo11} and \cite{Pavlovski09}.

Depending on the line(s) included, the primary is found to contribute
68 to 72 percent of the total light, with hydrogen lines supporting the
larger fractions. Hydrogen suggests a few percent lower temperature for
the primary, compared to the starting values. This is compatible with
the tendency seen in Figure 7 of \cite{H2000}, that H and He
lines for the primary only marginally agree on effective temperature
(taking minimum $\chi^2$ at the relevant gravity, a 1000\,K difference
in temperature estimation occurs).

The inconsistency between different indicators underlines the importance
to develop a more consistent atmosphere model for these stars. One way,
following \cite{Nieva12}, is to include more ionization
equilibria by analyzing the full wavelength range covered by the new
spectra. This work-intensive analysis is out of the scope of the present
paper, but probably indispensable to constrain better the degeneracy in
the determination of the radii. Its success might be limited by the
rotational broadening in the spectra. Another point of attention is the
need to take into account temperature and gravity variations over the
surface, due to the slightly non-spherical shape of the stars. 
Our work shows that the purely photometrically estimated  
light factors (Table~\ref{Table:lf}) lie within the broader range  
$0.68 - 0.72$ of light factors (primary to total light) derived here   
from the {\sc hermes} spectra. However, there are some warnings 
that improvement is needed - the spectroscopic
estimates may be biased as different indicators are not yet fully
compatible.



\subsection{Light Curve Analysis}

We use EB modeling software {\sc phoebe} \citep{PrsaZwitter05} 
based on Wilson-Devinney code \citep{WD71,W79}
for our light curve analysis. We fit light curves spanning 
$40$ yrs, covering one full apsidal motion cycle and 
in Johnson $UBV$ and  Str\"{o}mgren $uvby$ photometry. 

\begin{figure}[ht]
\epsscale{1.0}
\begin{center}
\includegraphics{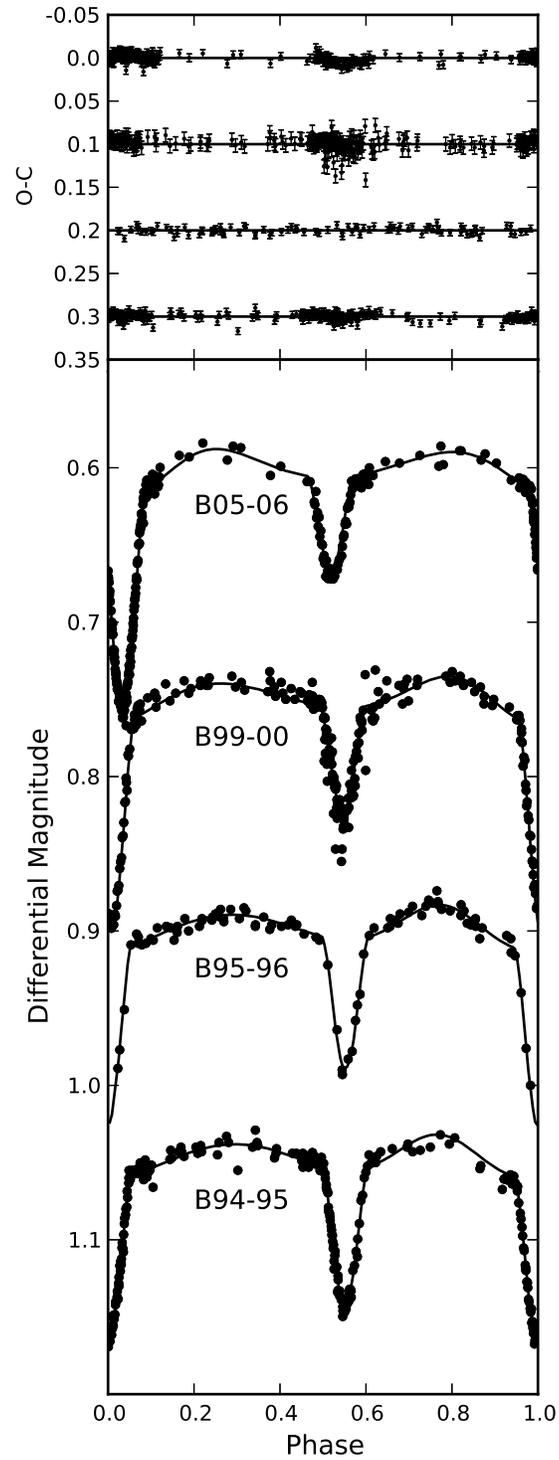}
\end{center}
\caption{\label{fig:B} Representative fits to light curves from 2005--2006, 
1999--2000, 1995--1996 and 1994--1995 in the Johnson $B$ passband from 
global fits to all light curve data, offset for clarity. 
The residuals to the fits $(O-C)$ are shown above. 
}
\end{figure}

\begin{figure}[ht]
\begin{center}
\includegraphics{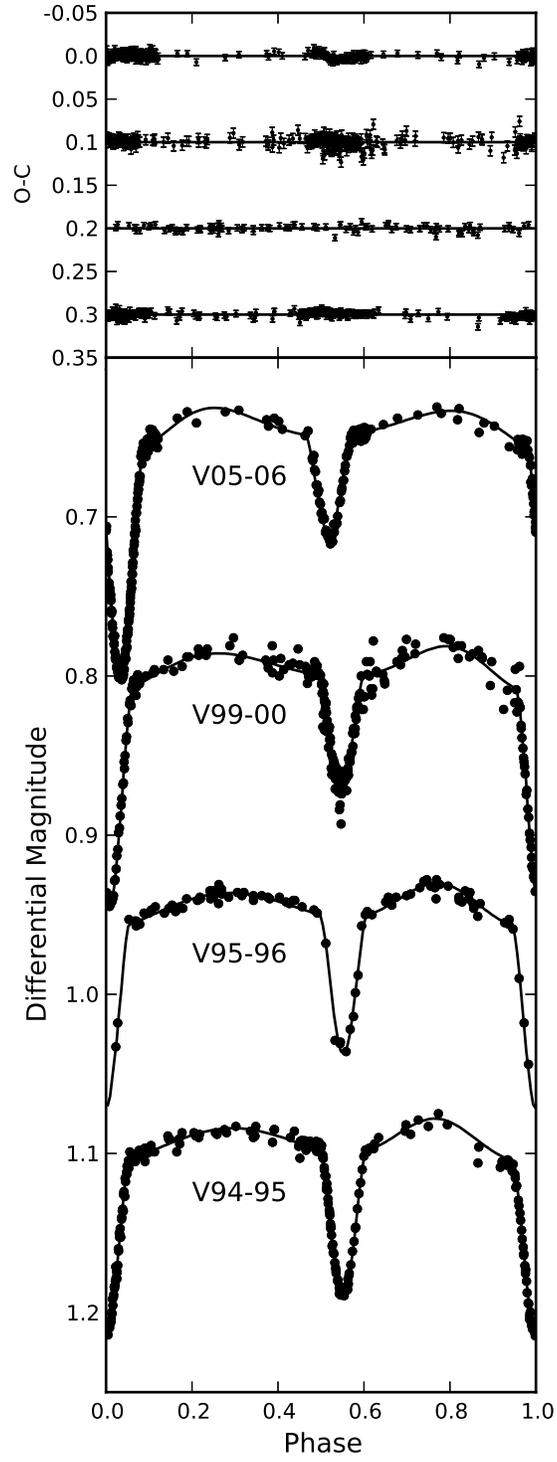}
\end{center}
\caption{\label{fig:V} 
Same as Fig.~\ref{fig:B}, but showing Johnson $V$ band light curves and fits.
}
\end{figure}

\begin{figure}[ht]
\begin{center}
\includegraphics{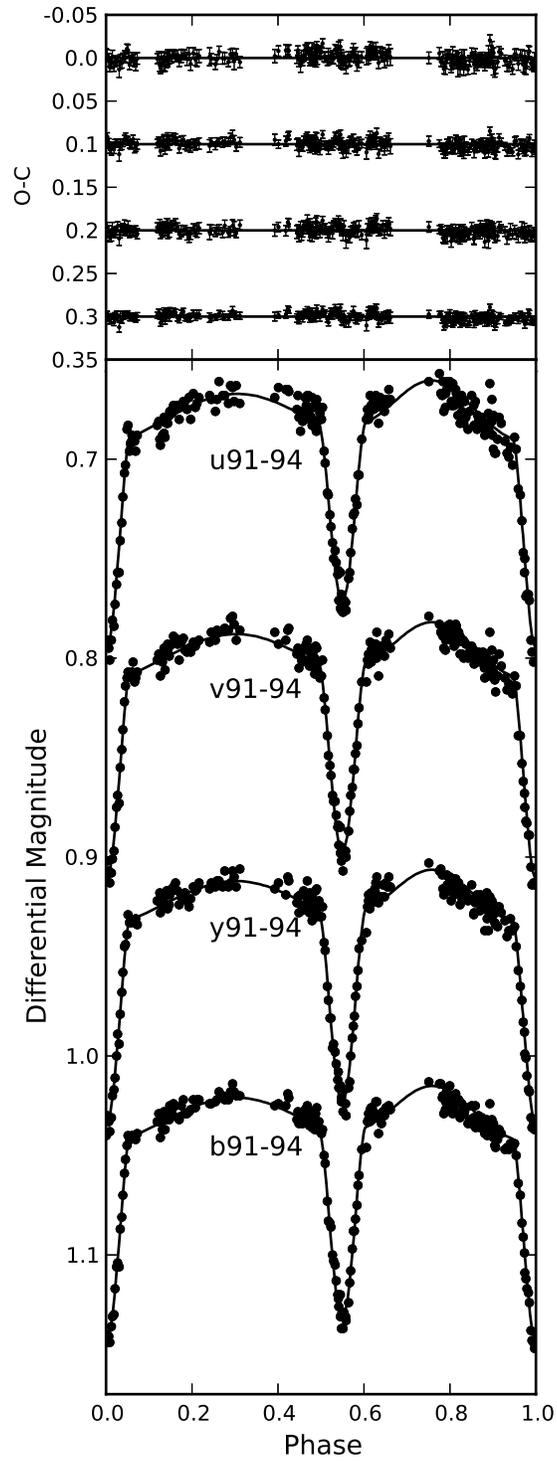}
\end{center}
\caption{\label{fig:SAT} 
Same as Fig.~\ref{fig:B}, but showing Str\"omgren $uvby$ light curves and fits.
}
\end{figure}

\begin{figure}[ht]
\begin{center}
\includegraphics{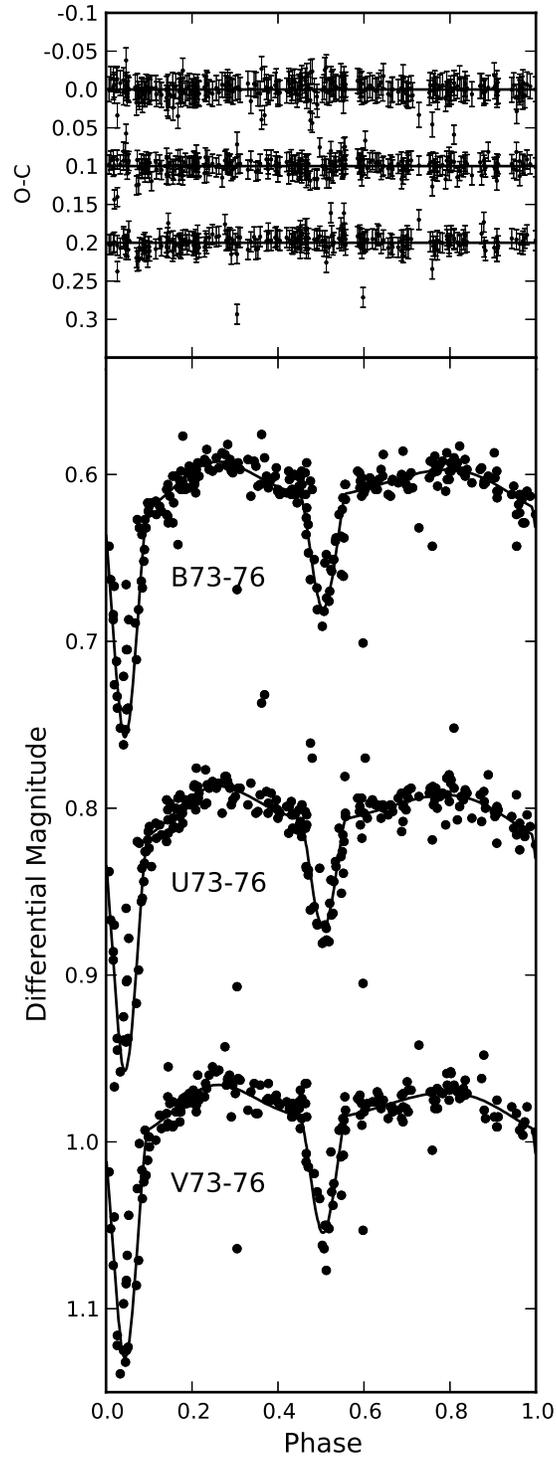}
\end{center}
\caption{\label{fig:KPNO} 
Same as Fig.~\ref{fig:B}, but showing 1973--1977 Johnson $UBV$ light 
curves and fits.
}
\end{figure}

Figures \ref{fig:B}, \ref{fig:V}, \ref{fig:SAT}, \ref{fig:KPNO} are
the residuals (data-model) for our global best fit model to the light curves 
for every light curve epoch 
and filter in Table~\ref{Table:Phot}. 
Overall, the residuals are small - typically $\approx0.005$ mag. The residuals are significantly 
larger for light curve epochs 1970-1984, since 
error bars on the photometry data points 
measured using photometric plates is larger. 
We explore ranges for our light curve parameters as listed in Table~\ref{table:grid}.
Our global best fit matches observations well - the final 
light curve parameters  
$\Omega_1$, $\Omega_2$, $i$, 
and $\frac{T_2}{T_1}$ are listed in
Table~\ref{Table:lc}.




\subsubsection{Setup \label{subsec:setup}}

For our global best fit light curve model, we adopt 
a square root limb darkening law \citep{Claret00}, 
a B1V spectral type for the primary star implying 
$T_1=30000$ K \citep{H2000}, no light reflection, 
and no third light.  

We have four light curve parameters 
of interest - primary potential
$\Omega_1$, secondary potential $\Omega_2$,
inclination $i$ and temperature ratio 
$\frac{T_2}{T_1}$. 
A ``parameter of interest'' is defined as
a parameter that is varied to
compute our confidence intervals.
We determine these parameters and their uncertainties by
mapping $\chi^{2}$ space. 
Potential $\Omega$ is a modified Kopal potential 
for asynchronous, eccentric orbits \citep{W79}. The potential 
($\Omega \propto R^{-1}$) takes into account contributions 
from the star itself, its companion, the star's rotation about its axis, and the 
star's rotation in its orbit. 

Our fixed parameters are the argument of periastron $w_0$, 
eccentricity $e$, apsidal motion $\dot\omega$, 
semi-major axis $a$, mass ratio $q$, period $P$, 
ephemeris HJD$_0$, systemic velocity $\gamma$, 
gravity brightening coefficients $g_1$ \& $g_2$,  primary and secondary 
synchronicity parameters $F_1$, $F_2$ and albedos $A_1$, $A_2$. 
We fix the argument of periastron $w_0$, eccentricity $e$ and apsidal motion $\dot\omega$ to 
values determined by a multi-epoch light curve analysis from \cite{Garcia11}. 
We fix mass ratio $q\equiv \frac{M_2}{M_1}$, semi-major axis $a$, 
orbital period $P$, time of minima HJD$_0$ 
and systemic velocity $\gamma$ to values from 
\cite{H2000} analysis of the spectroscopic orbit. As mentioned previously, 
our {\sc hermes} spectra analysis derives radial velocity amplitudes 
$K_1$ and $K_2$ in agreement with the \cite{H2000} 
spectroscopic orbit (see Table~\ref{Table:RV}). 
We adopt gravity brightening coefficients ($g_1, g_2$) and surface albedos ($A_1, A_2$) 
to be $1.0$ as appropriate for stars with radiative envelopes. 
The gravity brightening coefficient $g_1=g_2=1.0$ for stars with radiative envelopes 
was first found by \cite{vonZeipel24}. 
We fixed rotational synchronicity parameters $F_1 = 1.13$ 
and $F_2 = 1.11$ to values from \cite{H2000}. 
Our limb darkening coefficients follow the square-root 
law for hot stars \citep{Claret00} and are listed in Table \ref{Table:ld}.






\subsubsection{Fitting Method \label{subsec:method}}

Our fitting method is adapted from G\'omez Maqueo Chew et al. (2014, in prep). 
We determine our best fit global light curve 
solution by finding a unique set of light curve 
parameters $\Omega_1$, $\Omega_2$, $\frac{T_2}{T_1}$ and $i$ that correspond to the
 minimum chi square $\chi^{2}_{\rm min}$ in 
 a well mapped grid of parameter space. 
 The chi square is a function of the light curve parameters, 
 $\chi^{2}=\chi^{2}(\Omega_1,\Omega_2,\frac{T_2}{T_1},i)$. 
We map parameter space by computing $\chi^{2}$ for a grid of $>10^{5}$ unique sets 
of these light curve parameters. 
We use our map of parameter space to compute the uncertainties on our 
light curve parameters using confidence intervals.  
Plots of $\Delta\chi^2$ vs stellar radii $R_1$, $R_2$, 
temperature ratio $\frac{T_2}{T_1}$, and inclination $i$ with confidence intervals
are shown in Figure~\ref{fig:degen}. 


The step-by-step procedure is as follows: 

1. We sample a coarse grid of $10^{4}$ points defined by a range 
of potential $\Omega_1$, potential $\Omega_2$, inclination $i$ and temperature ratio 
$\frac{T_2}{T_1}$. The parameter ranges and spacings are given in Table~\ref{table:grid}. 

For each grid point, we fit only for the ``light levels" in {\sc phoebe} 
which is equivalent to the total light contribution from each star in
the photometric bandpass. 
We avoid using the WD2003 differential corrections (DC) 
fitting algorithm within {\sc phoebe} to fit 
our light curve parameters. 
The DC algorithm can fall into local 
minima when fitting for many parameters. 
We compute the total chi square $\chi^{2}_{k}$ 
for each light curve fit as the sum of the chi square 
$\chi^{2}_{p}$ at each passband and epoch: 
\begin{equation}
\chi^{2}_{k}(\Omega_{1k},\Omega_{2k},\frac{T_{2k}}{T_1},i_{k})=\sum^{15}_{p}\frac{\chi^{2}_{p}}{\sigma_p^{2}}
\end{equation}
Where index $k$ corresponds to a unique 
point in parameter space ($\Omega_{1k}$, $\Omega_{2k}$, $\frac{T_{2k}}{T_1}$, $i_{k}$). 
$\chi^{2}_{k}$ is the total chi square over 
all light curves at a unique point $k$. 
Index $p$ corresponds to a unique light 
curve passband epoch as specified in Table~\ref{Table:Phot}. 
The chi square at specific passband $\chi^{2}_{p}$ is computed as: 
\begin{equation}
\chi^{2}_{p} = \sum^{N}_{i}\frac{(f -f_{m})^{2}}{\sigma_i^{2}}
\end{equation}
Where $N= N_d - N_p = 3485$ is the number of photometry data points $N_d$ minus the 
number of parameters of interest $N_p$ over all light curve 
epochs. Each data point has an error bar $\sigma_i$. 
Each light curve at a specific epoch and filter 
has a multiplicative factor $\sigma_p$ which takes into account 
systematic error. Multiplicative factor $\sigma_p$ is 
used to normalize the $\chi^2$ such that 
$\chi^{2}_{\rm min}=N$ or reduced $\chi^{2}_{\rm min,red}=1.0$. 
$f$ is the total flux of the binary at an HJD, 
and flux $f_{m}$ is the corresponding model. 
From our coarse grid, we find the 
minimum total chi square $\chi^{2} = \chi^{2}_{\rm min}$ in parameter space. 

2. We adjust the error bars of the individual photometry data points 
for all light curves to take into account any systematic error. 
For the minimum $\chi^{2}_{\rm min}$ solution, the passband $\sigma_p$ is computed for 
each separate light curve epoch and filter using the equation: 
\begin{equation}
\sigma_p = \sqrt{\frac{N}{\chi^2_{\rm min}}}
\end{equation}
Where $N=3486$ as in step 1, and $\chi^{2}_{\rm min}$ is the minimum total 
$\chi^2$ of the coarse grid. 
We choose compute multiplicative factor $\sigma_p$ to weight each light curve such that 
the minimum reduced chi squared $\chi^{2}_{\rm min,red} = 1.0$ 
for our global best fit solution. 
We then rescale the $\chi^{2}$ of 
all other light curve fits using the passband $\sigma_p$: 
\begin{equation} 
\chi^{2}_{k} = \sum^{15}_{p}\frac{\chi^{2}}{\sigma_p^{2}}
\end{equation}
Where $\chi^{2}$ un-scaled and $\chi^{2}_{k}$ is the scaled 
chi square at a unique point in parameter space $k$. 



3. We perform steps 1 and 2 for a fine grid of $>10^5$ points in parameter space around the 
location of the minimum $\chi^2_{\rm min}$. 
In this way we carefully map out parameter space at the location of the 
$\chi^{2}_{\rm min}$. We use multiple fine grids to precisely find our global best 
fit minimum. The average grid spacings 
are \OmegaOneSpF, \OmegaTwoSpF,
\TfracSpF~and~\InclSpF~respectively for
$\Omega_1$, $\Omega_2$, $i$ and $\frac{T_2}{T_1}$. 

We find that the location of the minimum $\chi^{2}$ moves slightly, and
we recompute the multiplicative factor $\sigma_p$ for each light curve
 to account for this, again making  
$\chi^2_{\rm min,red} = 1.0$. Finally, we 
have a global best fit solution within a finely sampled 
parameter space. Our global best fit solution listed in 
Table~\ref{Table:lc} corresponds 
to the point in parameter space where 
scale chi square by $\sigma_p$ such that $\chi^{2}_{\rm min,red} = 1.0$.

\subsubsection{A Comparison of Light Curve Models\label{subsec:compmodels}}
 
In order to ensure our light curve solution 
is robust and thus our light curve parameters 
are accurate, we compare our best fit light curve model
described above with several other models.
As shown in Table~\ref{Table:Models} we find little 
effect on our best fit light curve parameters from using 
different light curve models. 
All other models 
are not as favorable due to larger $\chi^{2}$ or 
temperatures that do not agree with the analysis 
of the component spectra of V578~Mon from spectral 
disentangling of \cite{H2000}.

For all the tests described below, we start at our best fit 
solution, then fit all light curves 
in {\sc phoebe} for primary potential $\Omega_{1}$, 
secondary potential $\Omega_{2}$, temperature ratio 
$\frac{T_{2}}{T_{1}}$, and inclination $i$. 
 Our global best fit uses a fixed primary temperature T$_1=30000$~K, 
no light reflection and no third light. Furthermore, 
our global best fit uses fixed square root law 
limb darkening coefficients, which are found to 
work best for hot ($T_{\rm eff}>9000$K) stars 
\citep{Diaz92,VanHamme93}. We discuss the different 
light curve models in the order in which they 
appear in our summary Table~\ref{Table:Models}: 
 \begin{enumerate}

\item {\bf Fitting for Limb Darkening Coefficients.} 
We test the effect of fitting for square 
root law limb darkening coefficients, 
finding a lower chi square due to a 
larger number of free parameters. We find 
little effect on $\Omega_1$, $\Omega_2$ or $i$. 
However - we do find a much lower 
$T_2=$\ldTtwo. We reject this light curve 
model since $T_2=$\ldTtwo~is significantly outside
of the acceptable range for $T_2=$\HensTtwo~from 
the spectral disentangling of \cite{H2000}. 
We therefore perform another test: we keep 
$\frac{T_2}{T_1}$ fixed to our best fit value, and fit for 
the limb darkening parameters, $\Omega_1$, $\Omega_2$
and $i$. We again find little effect on $\Omega_1$, 
$\Omega_2$ or $i$.

\item {\bf Using a different Limb Darkening Law.} 
We test the linear cosine and logarithmic limb darkening laws, 
finding little effect on our light curve parameters. The linear cosine 
law has a lower $\chi^{2}=$\coschisqr~than our best fit model $\chi^{2}=$\bestchisqr. 
The light curve model with logarithmic limb darkening 
has a larger $\chi^{2}=$\logchisqr~-~we therefore reject this model. 
See Table \ref{Table:ld} for a list of the theoretical limb darkening 
coefficients for each light curve model that we test. 

\item {\bf Changing the assumed Primary Star Temperature.} 
We test the effect of changing our adopted primary star effective 
temperature $T_1$. Our adopted primary temperature for our best fit solution 
is $T_1=$\bestTone~K.  Once again, we find little effect on $\Omega_1$, 
$\Omega_2$, $i$ or $\frac{T_2}{T_1}$. 

We start with our best fit global solution, but 
set $T_1 = 31500$~K and $T_1 = 28500$~K, $3\sigma$ above and 
below our adopted primary star effective temperature.
Fits with lower primary temperature $T_1$ do result in a better $\chi^{2}$ - 
however, $T_1 < 29000$~K does not agree with the spectral disentangling 
analysis from \cite{H2000}. This may be due to the fact that the {\sc phoebe} 
light curve analysis constrains the temperature ratio and 
not the individual temperatures themselves. Further light curve 
tests at lower preferred temperatures $T_1$ and $T_2$ confirmed 
that changing effective temperatures have little effect on the 
geometric parameters, $\Omega_1$, $\Omega_2$ and $i$.

\item {\bf Light Reflection.} We fit our light curve model with one light reflection. 
We find an inclination $i$ larger by $2\sigma$. However, 
the $\chi^{2} =$\refchisqr~is higher than our best fit $\chi^{2}=$\bestchisqr. 
We reject this model on this basis. 

\item {\bf Third Light.} We test the possibility of
 third light and its effect on our 
best fit parameters. We fit for a third light parameter 
starting from our best fit light curve solution.
The third light model has a lower 
$\chi^2$ due to a larger number of free parameters. We find 
$\Omega_1$ and $i$ to be larger by $2\sigma$ and $2.5\sigma$ 
respectively from our best fit model. 

However, the third light parameter $L_3$ 
varies on the order of an apsidal period of the system. As shown in Table \ref{Table:l3}, 
we find at max a small contribution of third light
$\frac{L_3}{L_{\rm tot}}\approx0.045$ for 
Johnson B filter of light curve epochs  
1967-1984 and 2005-2006. 
This is likely due to {\sc phoebe} using the $L_{3}$ parameter to 
minimize the small systematic error of $0.005$ mag in the residuals 
of the 1967-1984 and 2005-2006 light curve epochs. 
Furthermore, the systemic velocity measured with the {\sc hermes} spectra
and the {\sc caspec} spectra in \cite{H2000}  does not give any 
evidence for a large third body in the system that would contribute 
significantly to the light. This is consistent with the third-light tests 
performed here. 

\end{enumerate}

\subsubsection{Uncertainties on Light Curve Parameters} 

\begin{figure}[ht]
\begin{center}
\includegraphics[angle=90,width=\textwidth]{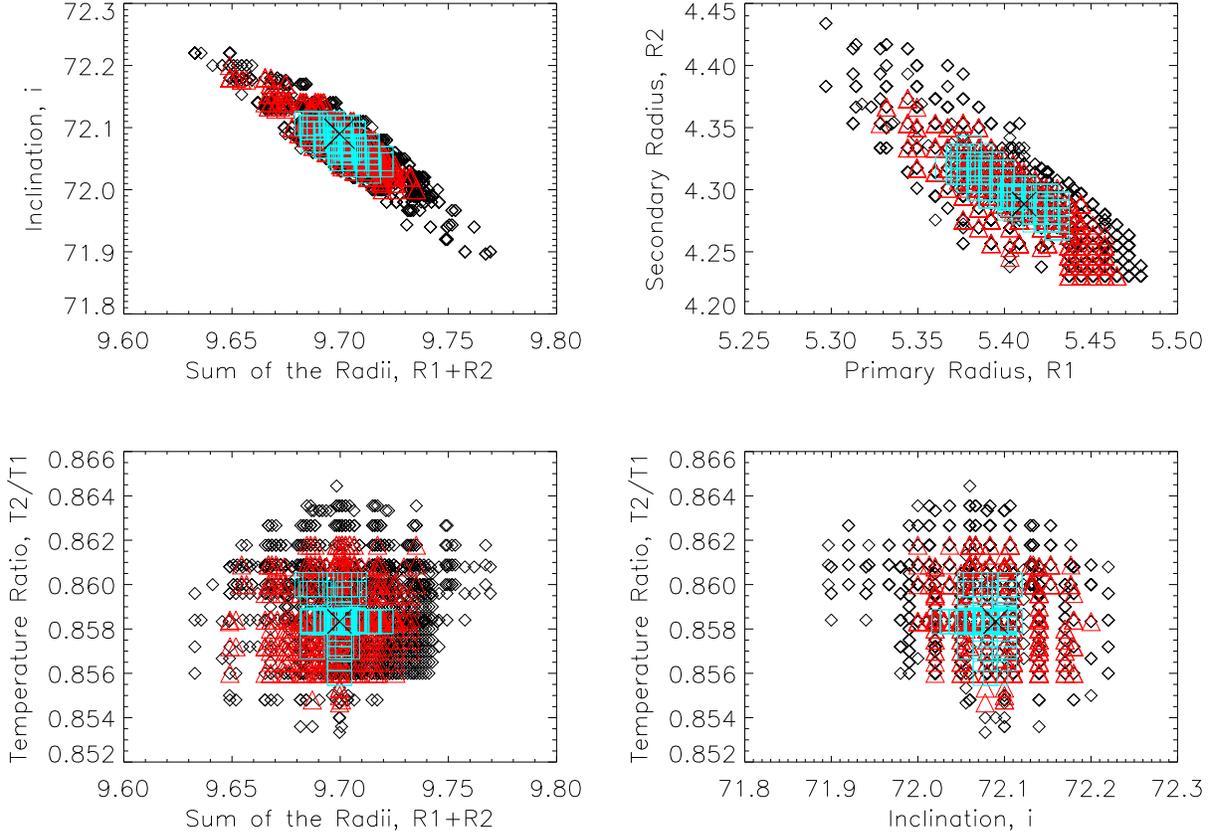}
\end{center}
\caption{\label{fig:degen} 
Degeneracies for our best fit light curve solution. The blue squares, red triangles and black diamonds 
correspond to difference in chi square from the global best fit solution  
$\Delta\chi^{2}=4.72$,$9.70$, and $16.3$ respectively. For four parameters of interest 
these $\Delta\chi^{2}$ correspond to $1\sigma$, $2\sigma$ and $3\sigma$ respectively. 
There is a small degeneracy between the sum of the radii $R_1+R_2$ and $i$. 
This degeneracy is typical for detached eclipsing binaries with circular or near circular orbits. Similarly, 
there is a small degeneracy between the 
primary and secondary radii $R_1$ and $R_2$. The global best fit solution 
is marked with an X. There is no degeneracy between the 
temperature ratio $\frac{T_2}{T_1}$ and inclination $i$ or sum of the radii 
$R_1+R_2$. A color version of this figure is available in the online journal. 
}
\end{figure}

We compute uncertainties on each parameter of interest using 
confidence intervals as shown in Figure~\ref{fig:degen}. 
From \cite{NR}, for four parameters of interest,
we find that $1\sigma$, $2\sigma$, and $3\sigma$ 
uncertainties  correspond to
solutions with confidence intervals of 
$\Delta\chi^2=\chi^2-\chi^2_{\rm min,red} = 4.72$, 
$9.70$ and $16.3$ respectively. Here, $\chi^{2}_{\rm min}$ is the 
minimum $\chi^{2}$ of our global best fit solution. 

From Figure~\ref{fig:degen} 
we see small degeneracies between the geometric parameters -
radii $R_1$, $R_2$, and inclination $i$. 
However - as expected we do not see degeneracies 
between the geometric light curve parameters and the temperature 
ratio $\frac{T_2}{T_1}$. 

Since $\frac{T_2}{T_1}$ is not strongly degenerate with these other 
parameters, we could potentially decrease the number 
of parameters of interest and in turn decrease the formal 
parameter uncertainties. Therefore, the uncertainties presented 
here are possibly conservative, given that we assume 
all degrees of freedom are parameters of interest \citep{Avni1976}.

The small degeneracies in our parameters leads to 
 uncertainties on potentials $\Omega_1$ and $\Omega_2$ 
of less than $<1.5$\% error - this error already takes 
into account any systematic error in fitting the light curves, as 
detailed in \S\ref{subsec:method}. Similarly, the uncertainty on 
the temperature ratio $\frac{T_2}{T_1}$ and inclination are also 
$<1$\%. 

A source of systematic uncertainty unaccounted for from the confidence 
intervals and fitting procedure in \S\ref{subsec:method} is from the
comparison of light curve models detailed in \S\ref{subsec:compmodels} 
and Table~\ref{Table:Models}. As shown in Table~\ref{Table:Models}, 
all other light curve models assessed in \S\ref{subsec:compmodels}
with the exception of using linear cosine LD parameters are not as 
favorable as our best fit model. The linear cosine model has a lower $\chi^{2}$. 
Nevertheless, The inclination $i$, 
temperature ratio $\frac{T_2}{T_1}$ and secondary potential $\Omega_2$ 
are all within $1\sigma$ of our best fit model. 
However, the primary potential for the linear cosine model 
$\Omega_1 = $\cosOone~as compared to our best fit 
$\Omega_1 = $\bestOone. Therefore our uncertainty on $\Omega_1$ from 
our best fit model could be slightly underestimated from these model comparisons.

\subsubsection{Consistency of Light Fractions\label{subsec:lf}} 
As mentioned by \cite{Torres10} an important consistency check of our 
light curve solution is that the light fraction $l_{f,1} = \frac{l_1}{l_1+l_2}$
determined from spectroscopy and photometry agree. Given the small 
degeneracy between $R_1$ and $R_2$ as seen in Figure~\ref{fig:degen}, we 
compare our photometrically determined light fraction
with the light fraction from the {\sc HERMES} spectral disentangling 
and a previous combined light curve and spectral disentangling analysis
 from \cite{H2000}. We find that all three light 
fractions agree with each other to within 1.2$\sigma$. A comparison of 
light fractions is shown in Table~\ref{Table:lf}. 

\begin{figure}[ht]
\begin{center}
\includegraphics[angle=90,width=\textwidth]{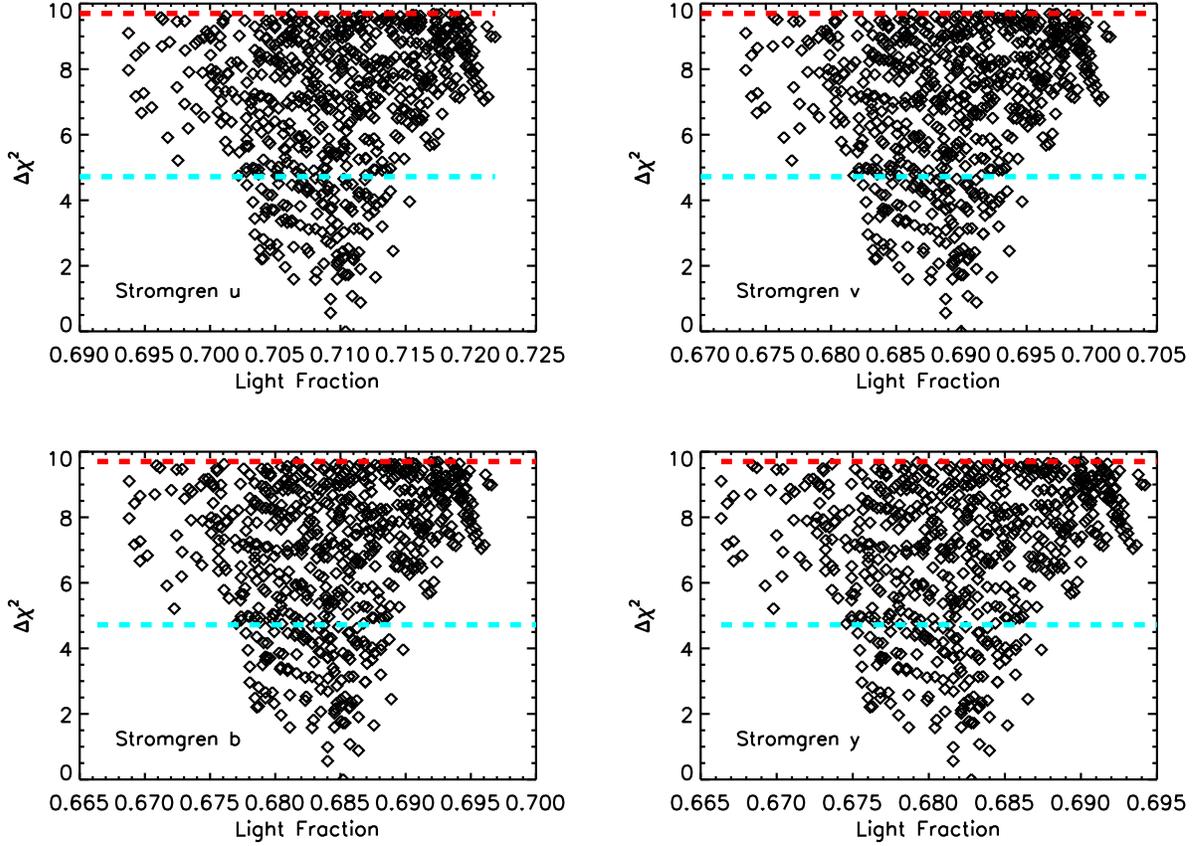}
\end{center}
\caption{\label{fig:lf1} 
The light fractions $l_{f,1}=\frac{l_1(\lambda)}{l_1(\lambda)+l_2(\lambda)}$ for light curve fits 
within $1\sigma$ (below the blue line) and $2\sigma$ (below the red line) uncertainty for the 
Stromgren $uvby$ photometry. Our light fractions are consistent with the light fractions computed from 
\cite{H2000}. A color version of this figure is available in the online journal. 
}
\end{figure}

\begin{figure}[ht]
\begin{center}
\includegraphics[angle=90,width=\textwidth]{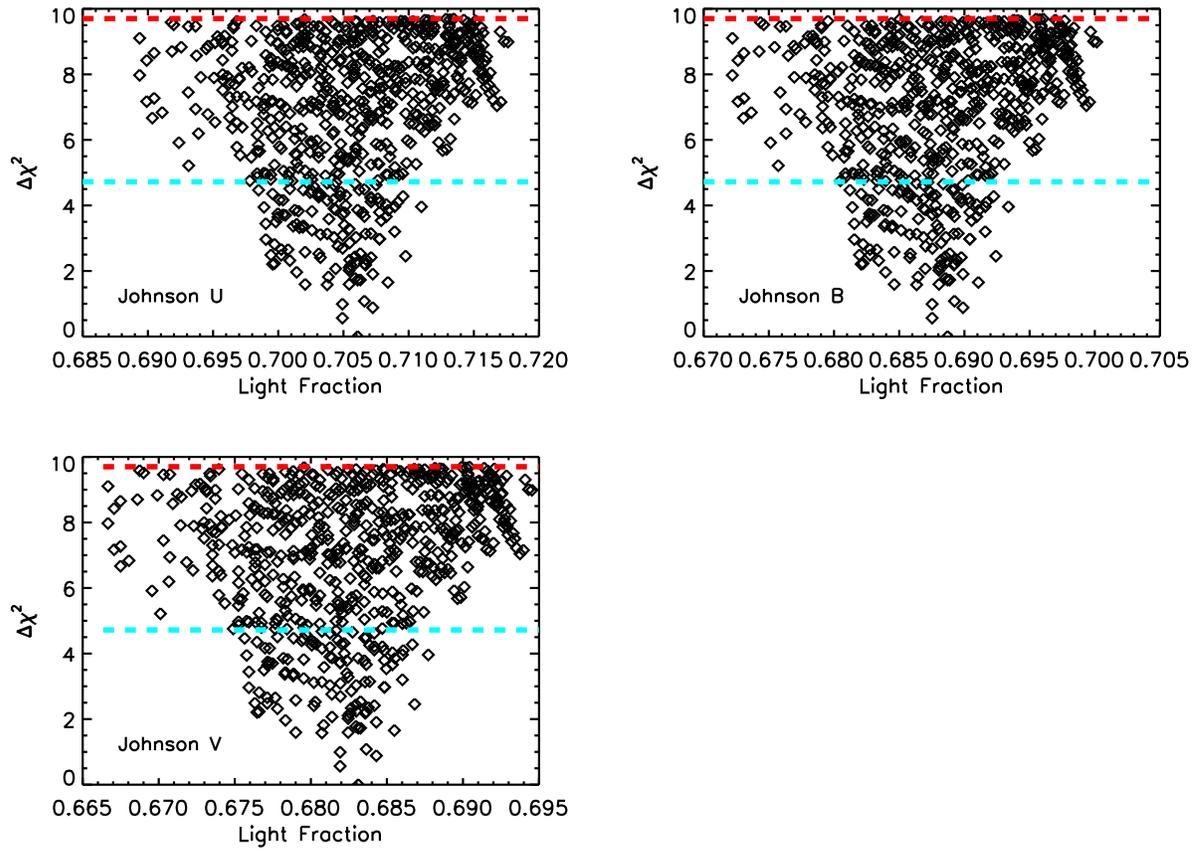}
\end{center}
\caption{\label{fig:lf2} 
Same as Figure \ref{fig:lf1} except for the Johnson UBV photometry. 
}
\end{figure}

For each of the $\approx10^{5}$ 
light curve fits to our 40 yrs of photometry data, we compute the light fraction 
at each of the passbands Johnson $UBV$ and Str\"{o}mgren $uvby$ photometry, 
$l_{f,1}(\lambda)=\frac{l_1(\lambda)}{l_1(\lambda)+l_2(\lambda)}$ 
where $l_1(\lambda)$ and $l_2(\lambda)$ are the contribution of
the primary and secondary star to the total light 
at a specific passband out of eclipse. The distribution 
of light fractions $l_{f,1}$ for light curve models with confidence intervals 
of $1\sigma$ and $2\sigma$ are shown in Figures \ref{fig:lf1} and \ref{fig:lf2}. 
 
\subsection{Comparison with \cite{H2000}}

\cite{H2000} uses an iterative, combined light curve 
and spectral disentangling analysis using the Wilson-Devinney light curve modeling 
program to compute their light cure parameters. We find that $R_1=$\HensRone~\Rsol~from 
\cite{H2000} is $2.5\sigma$ discrepant from our best fit $R_1=$\bestRone~\Rsol. We find that our 
inclination $i=$\bestIncl~deg is 1.6$\sigma$ discrepant from 
$i=$\HensIncl~deg from \cite{H2000}. These discrepancies  are likely due 
to the addition of apsidal motion and an updated eccentricity determined 
in \cite{Garcia11}. Apsidal motion and eccentricity 
can affect the potentials $\Omega_1$ and $\Omega_2$ and hence the determination 
of the radii at a low level. The potential $\Omega$ for a non-circular orbit 
is a function of eccentricity (see \cite{W79}). 
The addition of more light curve epochs may also play a role. 
\cite{H2000} only uses the 1991-1994 light curve epoch with Str\"{o}mgren $uvby$ photometry. 
As a check, we also recover the \cite{H2000} light curve solution when 
we fit only the 1991-1994 light curve epoch. Finally, simply the addition of more 
photometry data points may play a role. We use 3489 photometry data points 
in our light curve solution, whereas \cite{H2000} use 992. Our best fit 
secondary radius $R_2=$\bestRtwo~\Rsol~is in agreement with 
\HensRtwo~\Rsol~from \cite{H2000}.  
Our best fit temperature ratio $\frac{T_2}{T_1}=$\bestTrat~is in agreement 
with the temperature ratio of~\HensTrat~from an analysis of the disentangled 
component spectra \citep{H2000}. 


\section{Results: Absolute Dimensions and Apsidal Motion of V578~Mon \label{sec:results}} 

The absolute dimensions and other fundamental 
properties of V578~Mon are compiled in Table~\ref{Table:absdim}. 
Here we detail how each fundamental parameter for V578~Mon is compute
in order of which they appear 
in Table~\ref{Table:absdim}: 
\begin{enumerate}

\item {\bf Orbital Period.} We adopt the orbital period $P=2.4084822$ d from \cite{H2000}. 

\item {\bf Masses.} The component masses M$_1=$\bestMone~\Msol\ 
and M$_2=$\bestMtwo~\Msol\ are determined 
from the spectroscopic orbit analysis from \cite{H2000}. 
We do not use RVs from our {\sc hermes} spectroscopy 
because the {\sc caspec} spectra have higher S/N - however, 
our analysis of the {\sc hermes} 
spectroscopy reconfirm the spectroscopic orbit. 

\item {\bf Radii.} We find precise uncertainties of $<1.5$\% 
for the primary radius $R_1=$\bestRone~\Rsol~and secondary 
radius  $R_2=$\bestRtwo~\Rsol~from our confidence intervals in 
Figure \ref{fig:degen}. 


\item {\bf Temperatures.} We find a $0.3$\% error on our temperature ratio $\frac{T_2}{T_1}=$\bestTrat~from 
our confidence intervals. Combined with 
the adopted temperature of the primary star T$_1=$\bestTone~K~\citep{H2000},
our temperature ratio of $\frac{T_2}{T_1}$
yields a secondary temperature of~T$_2=$\bestTtwo~K via propagation of errors.

\item {\bf Rotational Velocities.} We compute surface rotational velocities 
of v$_{\rm 1,rot}=$\bestvone~km s$^{-1}$ and 
v$_{\rm 2,rot}=$\bestvtwo~km s$^{-1}$ using the observed projected surface 
velocities v$_1\sin{i}=$\Hensvonesini~km~s$^{-1}$~and 
v$_2\sin{i}=$\Hensvtwosini~km~s$^{-1}$~from \cite{H2000} and our
inclination of i$=$\bestIncl. The uncertainty 
on rotational velocities are computed from propagating the error on 
the inclination $i$ and the observed v$\sin{i}$. 

\item {\bf Surface Gravities.} We compute the surface gravity $\log{g}$ from our masses and radii, finding 
$\log{g_1}=$\bestloggone~cm~s$^{-2}$ and $\log{g_2}=$\bestloggtwo~cm~s$^{-2}$. 
We compute the uncertainty on $\log{g}$ via error propagation: 
\begin{equation} 
\sigma_{\log{g}} = \sqrt{(\frac{\sigma_{M}}{M \ln{10}})^{2} + (\frac{2\sigma_{R}}{R\ln{10}})^{2}}
\end{equation}
Where $\sigma_{M}$ is the uncertainty on the mass and $\sigma_{R}$ is the uncertainty 
on the radius. 

\item {\bf Luminosities.} From our radii and temperatures, we compute compute luminosities for the primary 
and secondary star of $\log{\frac{L_1}{\Lsol}}=$\bestLone~
and $\log{\frac{L_2}{\Lsol}}=$\bestLtwo. We compute the uncertainty on the luminosity
using a similar error propagation as above, using errors from the temperature 
and radii, $\sigma_{T}$ and $\sigma_{R}$. 

\item {\bf Synchronicity Parameters.} We find the components of V578~Mon to be close but not exactly tidally locked, 
with $F_1=$\bestFone~and~$F_2=$\bestFtwo. The synchronicity parameter $F = \frac{w}{w_{\rm orb}}$, where $w$ is the rotational velocity at the surface $v_{\rm rot}$ and $w_{\rm orb} = \frac{2\pi R}{P}$ is the synchronous velocity.
We compute the uncertainty via propagation of error from $\sigma_R$, error on inclination 
$\sigma_i$, and error on projected rotational velocities $\sigma_{v\sin{i}}$. 

\item {\bf Internal Structure Constant.} One of us (Dr. Claret) computes the newtonian and general relativistic contributions to the observed internal structure constant, $\log{k_{\rm 2,newt}}=$\bestlogktwonewt and $\log{k_{\rm 2,GR}}=$\bestlogktwogr. 

\end{enumerate}




\section{The Stellar Evolution Models and Tests} 

We compare the absolute dimensions of V578~Mon to 
the stellar evolution models of three separate groups: 
(1) Geneva models of \cite{Georgy13} and \cite{Ekstrom12} hereafter Geneva13;  
(2) Utrecht models of \cite{Brott11} here after Utrecht11 
\footnote{The Utrecht Stellar Evolution group is now located in Bonn, Germany}; and
(3) Granada models of \cite{Claret04,Claret06} hereafter Granada04. We assume that both 
stars have the same initial chemical composition and age, as expected for tight binary systems. 
We perform two tests: (1) The  ``isochrone test'' , which tests the ability 
of stellar evolution models to produce stars with different masses, radii, temperatures, 
rotational velocities, and surface compositions at the same age;
and (2) The ``apsidal motion test'' which tests the ability of the stellar 
evolution models to reproduce the observed internal 
structure constant $\log{k_2}$ as determined from the observed 
apsidal motion. 

A comparison of the basic input physics of the models is 
given in Table~\ref{Table:evol}.  The models use the same opacity tables of 
\cite{IglesiasRogers96}. The mixing length 
$\alpha_{\rm MLT}\equiv\frac{l}{H_p}$ for all three sets of models 
differ by only $0.18$ at maximum. The stellar evolution models use similar mass 
loss treatment from the prescription by \cite{Vink01}. Given the probable 
young age of V578~Mon due to its location in the open cluster 
NGC 2244 of the Rosette Nebula, the components of V578~Mon 
are not expected to have undergone significant mass loss \citep{Vink01}. 

However, all three sets of models differ on the choice of the 
convective core overshoot parameter $\alpha_{\rm ov}$. 
For the H and He burning phases of the convective core, the convective 
core size of the star is enlarged by $R_{\rm cc} = R_{cc}(1+\frac{d_{\rm over}}{H_p})$, where 
$\alpha_{\rm ov}\equiv\frac{d_{\rm over}}{H_p}$ in units of pressure scale height. 
The overshoot parameter is designed to accomodate
for the non zero velocity of the material moving from the convective core to radiative zone 
of the star. Observationally, a larger overshoot parameter means longer MS 
lifetimes for a given star, and thus older ages.  
The Geneva13 models use a small convective core overshoot of $\alpha_{\rm ov}=0.1$ calibrated 
on width of the main sequence for stars with masses $M = 1.35-9.0~\Msol$
which is characterized by the red most point on the B-V, $M_{V}$ 
HR diagram (see figure 8 of \cite{Ekstrom12}). 
The width of the main sequence is defined theoretically by the 
end of the hydrogen burning phase.  
The Utrecht11 models use a high convective core overshoot of $\alpha_{\rm ov}=0.335$ which is
calibrated using the observed width of the main sequence from the VLT-FLAMES 
survey of B stars \citep{Evans05,Hunter07}. The convective 
core overshoot parameter $\alpha_{\rm ov}=0.335$ is chosen such that a 16 $\Msol$ star ends its MS 
lifetime when $\log{g} = 3.2$. This $\log{g}$ coincides with the drop in B star rotation rates 
in a $\log{g}$-$v\sin{i}$ diagram, which is interpreted as an 
estimate of the width of the main sequence for B stars. See \cite{Brott11} for an 
in depth discussion. The Granada04 models utilize a moderate 
convective core overshoot $\alpha_{\rm ov}=0.2$, though we performed 
several tests varying $\alpha_{\rm ov}$. 


Rotationally driven mixing can bring more 
H and He from the envelope to the core, thus extending the 
MS lifetime of the star - likewise a larger overshoot parameter extends 
the size of the core, leading to a longer MS lifetime.  The Granada04
models do not incorporate rotational mixing, while the 
Geneva13 and Utrecht11 models do. However -
All three sets of models include rotation.  All three sets of models 
use similar metallicity compositions of near solar. 
The initial bulk composition for V578~Mon is expected to 
be close to solar given that Mg surface abundance is within 
error of the solar surface abundance, 
despite the fact that several atmospheric 
abundances such as C, N and O are 
somewhat metal poor compared
to the Sun \citep{PavHens05}. This is because 
Mg abundance is not expected to be altered from 
the initial abundance in a star, where as C, N and O 
atmospheric abundances could vary in V578~Mon due to 
rotational mixing \citep{L2005}.  However, given that the 
C, N and O atmospheric abundances of V578 Mon may be 
lower than solar, the metallicity of V578 Mon 
still remains as a source of 
systematic error in comparing 
the evolution models to the observations.


The Granada04 models also compute the internal structure constants 
$\log{k_2},\log{k_3}$ and $\log{k_4}$ allowing for a test of 
the internal structure of V578~Mon via apsidal motion. Here 
we consider only the $k_2$ constant, given that $k_3$ and $k_4$ are very small. 
For V578~Mon, the tidal Love numbers 
quantify the deformation for each star's gravity field due to the companion.



\subsection{Isochrone Test for V578~Mon}

\begin{figure}[ht]
\begin{center}
\includegraphics[angle=90,width=\textwidth]{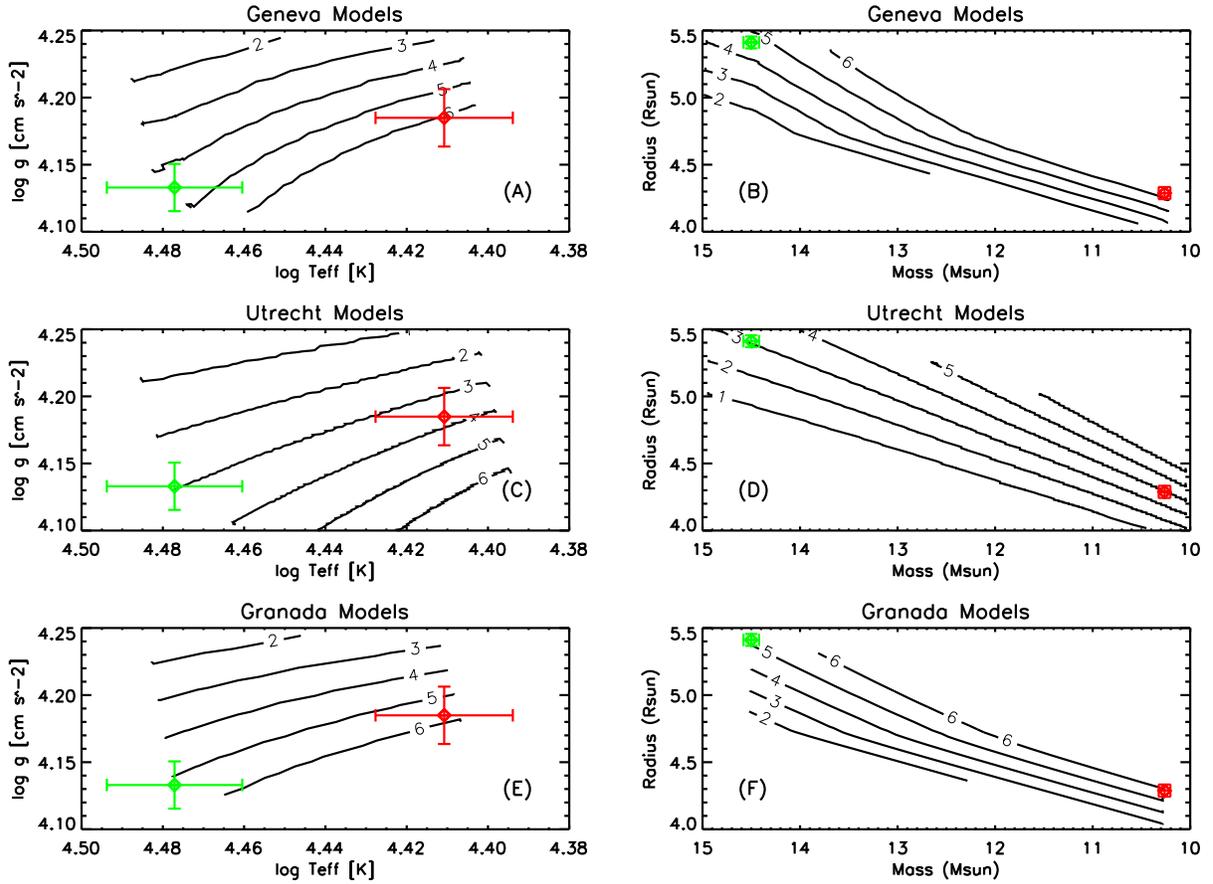}
\end{center}
\caption{\label{fig:iso} 
The best matches to observations are the Utrecht11 and 
Granada04 models, which both use larger than conventional overshoot of 
$\alpha_{\rm ov}=0.2$ - see Table~\ref{Table:ages} for details. 
Isochrones are in steps of 1 Myr of the Geneva13, Utrecht11 and Granada04 models. 
The green point is the primary star, and the red point is the secondary star. 
All models have rotational velocities that match the observed
velocities of V578~Mon $v_{\rm 1,rot}=$\bestvone~km s$^{-1}$ and 
$v_{\rm 2,rot}=$\bestvtwo~km s$^{-1}$.  A color version of this figure is available in the online journal. 
}
\end{figure}

In Figure~\ref{fig:iso}, we place the primary and secondary 
star on mass-radius and $\log{g}-\log{T_{\rm eff}}$ isochrones  
for each set of models. For the stellar evolution models 
to pass the ``isochrone test'' the models 
should predict a common age for both components of 
V578~Mon within uncertainty. Given how different the masses 
of the primary and secondary star for V578~Mon are the 
``isochrone test'' provides a stringent test of stellar evolution
models. We also match all evolution models to the rotational 
velocities of the primary and secondary star. 

We find several Geneva13, Utrecht11 and Granada04 models predict masses, 
radii and temperatures for the components of V578~Mon that 
fall within $1\sigma$ uncertainty of the
measured absolute dimensions. Therefore we estimate an age range for 
each star as shown in Table~\ref{Table:evol}. The age 
difference for Geneva13, Utrecht11 and Granada04 models is given
as the smallest possible difference between the ages of the two stars given 
age range of each star.

For the Geneva models we use isochrones with 
initial rotational velocities $\frac{v_i}{v_{\rm crit}}=0.30$ and
$\frac{v_i}{v_{\rm crit}}=0.35$ which allows us  
to match the observed rotational velocities for each star. We interpolate the 
model evolution tracks for the primary and secondary star using the online 
interactive tool provided by the Geneva group
\footnote{{\sf http://obswww.unige.ch/Recherche/evol/-Database-}} . Attempts to match 
the observed rotational velocities of V578~Mon with lower ($\frac{v_i}{v_{\rm crit}} < 0.30$) 
or higher ($\frac{v_i}{v_{\rm crit}} > 0.40$) initial velocities for either star were 
unsuccessful. Attempts to find a single initial rotational velocity to reproduce 
the current observed rotational velocities for both stars with reasonable predicted radii and masses 
were also unsuccessful. However, given that V578~Mon is very near
synchronization with the orbital period  ($F_1=$\bestFone, $F_2=$\bestFtwo), the 
rotational history of V578~Mon could be different from the best matched 
$\frac{v_i}{v_{\rm crit}}$ found here. If the initial velocities of the components of
V578~Mon were larger at the ZAMS than the orbital velocity, the stars could
spin down to synchronize with the orbital velocity. 
Conversely, if $\frac{v_i}{v_{\rm crit}}$ was smaller than the orbital 
velocities, then the components of V578~Mon could spin up \citep{Song13}. 
From Table \ref{Table:ages} we find an age difference of \mrGenevaDiff~Myr for mass-radius 
isochrones, and an age difference of only $\gtGenevaDiff$~Myr for $\log{g}-\log{T_{\rm eff}}$ isochrone. It is
easier to find consistency for the latter isochrones given our uncertainty in the effective temperatures 
of the two stars. We find that a primary radius of $R_1=5.50~\Rsol$ and 
a secondary star radius of $R_2=5.20~\Rsol$ yields common 
ages for the Geneva13 models. However, 
these radii are $3\sigma$ larger and $3\sigma$ smaller than our 
best fit model, respectively. 

For the Utrecht11 models we use isochrones that match the observed 
surface velocities of the components of V578~Mon, v$_{\rm 1,rot}=$\bestvone~km s$^{-1}$ and 
$v_{\rm 2,rot}=$\bestvtwo~km s$^{-1}$. The Utrecht11 models 
are computed at very small steps of mass and initial rotational velocity, 
such that interpolating between model tracks is unnecessary. 
From Table \ref{Table:ages} we a marginally common age 
(age difference $\mrUtrechtDiff$~Myr) for mass-radius 
isochrones, and a common age of $\gtUtrechtDiff$~Myr for $\log{g}-\log{T_{\rm eff}}$ isochrone. The 
models were computed at solar metallicity by Dr. Brott (private comm.). 

We compute the Granada04 models at 
the masses of primary and secondary star and chose rotational 
velocities to match the observed rotational velocities of V578 Mon. We attempt 
to match the absolute dimensions of V578~Mon to $\log{g}-\log{T_{\rm eff}}$ 
or alternatively mass-radius isochrones for V578~Mon. 
We find an age gap of 1.5 Myr for mass-radius 
isochrones, and a marginally common age for $\log{g}-\log{T_{\rm eff}}$ isochrones, when both 
stars have an overshoot of $\alpha_{\rm ov}=0.2$. Again - finding a 
match on the $\log{g}-\log{T_{\rm eff}}$ isochrones is easier given the greater 
uncertainty in the effective temperatures. 

\begin{figure}[ht]
\begin{center}
\includegraphics[angle=-90,width=\textwidth]{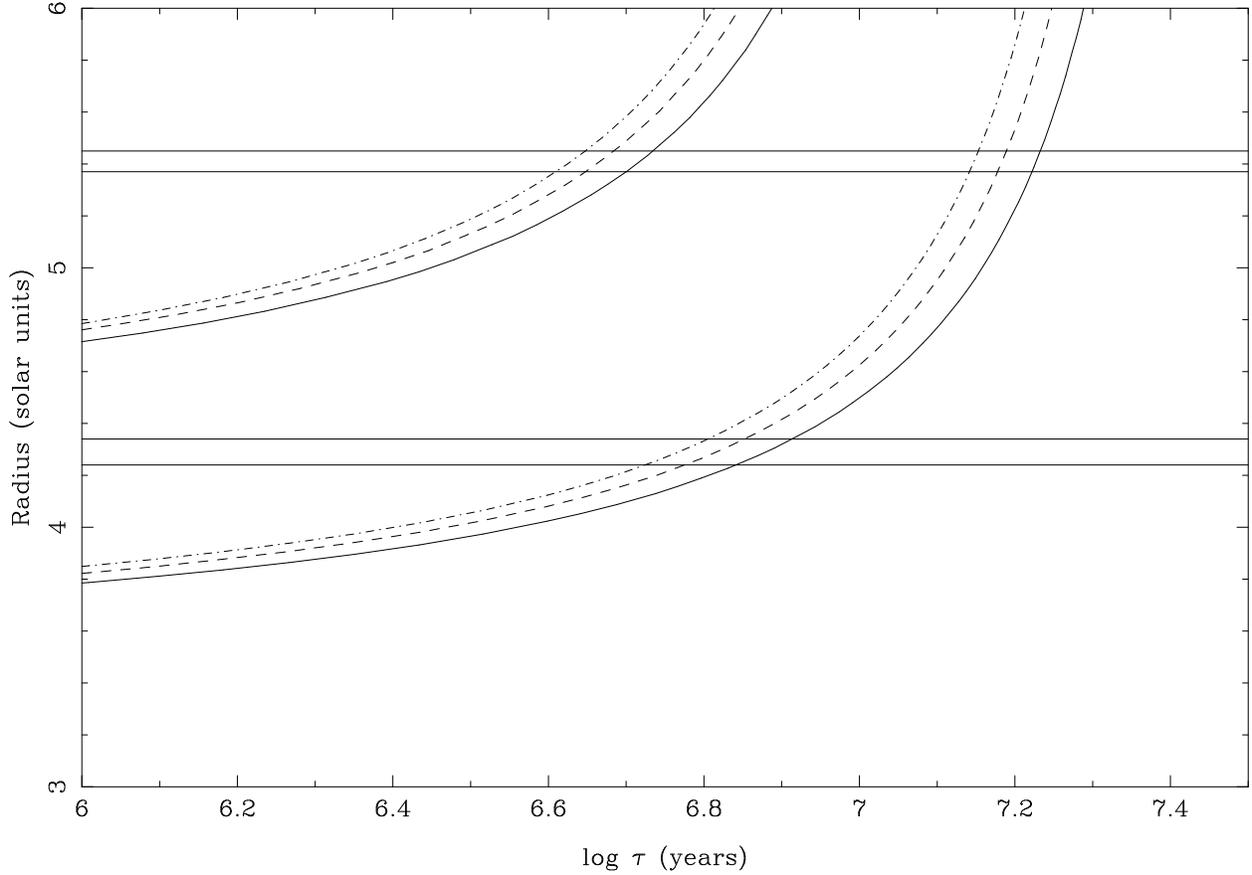}
\end{center}
\caption{\label{fig:diffovR} 
The time evolution of the radii for V578~Mon from Granada04 models 
computed for the masses of the V578~Mon primary and secondary. Dot-dashed, 
dashed, and solid lines are evolutionary models 
at convective overshoot of $\alpha_{\rm ov}$ of $0.2$, $0.4$ 
and $0.6$ pressure scale heights respectively. Horizontal lines are the upper and 
lower limit of uncertainty on the primary star and secondary star
radius respectively. The models predict a common age 
of 5.5 Myr - if we use a high convective overshoot of $\alpha_{\rm ov} = 0.6$ evolution model
for the primary star and $\alpha_{\rm ov}=0.2$ for the secondary star. 
}
\end{figure}

In an attempt to match the ages of the 
two stars on a mass-radius isochrone, we also compute Granada04 models for $\alpha_{\rm ov}=0.4$ and 
$\alpha_{\rm ov}=0.6$. Figure \ref{fig:diffovR} demonstrates the time evolution 
of the radii for V578 Mon for these different models. We find a near match 
on a single mass-radius isochrone with 
an age difference of only \mrGranadaDiff~Myr -  if we assume that the primary star 
has a convective overshoot $\alpha_{\rm ov}=0.6$
and the secondary star has a convective overshoot of $\alpha_{\rm ov}=0.2$. We also 
find a common age of $\gtGranadaDiff$~Myr for the $\log{g}-\log{T_{\rm eff}}$ isochrone. 
This does not mean that an $\alpha_{\rm ov}=0.6$ for the primary star
is correct for V578~Mon - merely that a higher convective overshoot 
allows for compatible ages between the two stars. High convective overshoot 
has been found to work in matching other EBs on a single isochrone 
\citep{Claret07}. 

In general, we find younger ages by $\approx1$ Myr 
for the Utrecht11 models of V578~Mon and similar ages 
for the Geneva13 and Granada04 models. This can be attributed to 
the larger convective overshoot of $\alpha_{\rm ov} = 0.355$ included in Utrecht11 models 
than in Geneva13 models ($\alpha_{\rm ov}=0.2$). While the primary star for the Granada04 
models does have an even higher convective overshoot of $\alpha_{\rm ov}=0.6$, the models 
do not include rotational mixing, which also extends the main sequence lifetime of the stars.  

\subsection{Apsidal Motion Test for V578 Mon} 
Measurement of apsidal motion in eccentric binary 
systems allow for a stringent test of 
the internal structure constant
$k_{\rm 2,theo}$ predicted from stellar evolution models
\citep[e.g.][]{Claret10}.  It is not possible 
to separate out each individual star's contribution to the apsidal 
period U from newtonian apsidal motion.

The apsidal motion for V578~Mon was measured by 
\cite{Garcia11}. The observed apsidal motion of the 
V578~Mon, \wdotactual, 
has contributions from both newtonian and general 
relativity components \citep{Claret10}:  
\begin{equation} 
\dot\omega_{\rm obs} = \dot\omega_{\rm newt} + \dot\omega_{\rm GR}  
\end{equation}
where $\dot\omega_{\rm GR}$ is given by 
\begin{equation} 
\dot\omega_{\rm GR} = 0.002286\frac{M_1+M_2}{a(1-e^{2})}
\end{equation}
We find that $\dot\omega_{\rm GR}=$\bestwdotgr~which is 
only $4\%$ of the newtonian apsidal motion 
$\dot\omega_{\rm newt}=$\bestwdotnewt. 

Both the newtonian and general relativistic observed apsidal 
motions $\dot\omega_{\rm newt}$ and $\dot\omega_{\rm GR}$ have associated observed
internal structure constants $k_{\rm 2,obs}$. The 
internal structure constant is twice the tidal love number \citep{Kramm11}, and 
is related to the density profiles, degree of sphericity, orbital parameters, masses, and
rotation rate of both components of a binary star. Specifically, the internal 
structure constant is related to the solution of the Radau differential equation as in equation
3 of \cite{Claret10}. Importantly - constant $k_{\rm 2,obs}$ is one 
the few ways to directly constrain the internal structure of stars. 

From the precise observed apsidal motion, 
we compute the observed internal 
structure constant, $k_{\rm 2,obs}$ = $k_{\rm 2,obs}(M_1, M_2, R_1, R_2, P, U, F_1, F_2, e)$, 
where U is the apsidal period, 
given by the equations (adopted from \cite{Claret10}): 
\begin{equation} 
k_{\rm 2,obs} = \frac{1}{c_{21} + c_{22}}\frac{P}{U}
\end{equation}
\begin{equation} 
c_{2i} = [(F_i)^{2}(1+\frac{M_{3-i}}{M_i})f(e)+15\frac{M_{3-i}}{M_i}g(e)](\frac{R_i}{a})^{5}
\end{equation}
\begin{equation} 
f(e)=(1-e^{2})^{-2}
\end{equation} 
\begin{equation} 
g(e) = \frac{(8+12e^{2}+e^{4})f(e)^{2.5}}{8}
\end{equation}
We compute the internal structure constant due to the newtonian apsidal motion, 
$\log{k_{\rm 2,newt}}=$\bestlogktwonewt, and due to general relativity, 
$\log{k_{\rm 2,GR}}=$\bestlogktwogr. The newtonian apsidal motion is 
much larger than the general relativistic component, and therefore 
the internal structure constant is also much larger. 

We compute the theoretical internal structure constant, 
$k_{\rm 2,theo}$ using the methods of \cite{Claret10}. The theoretical 
$k_2$ constant was corrected for by rotation \citep{Claret99} and 
dynamical tides \citep{WillemsClaret02}. The theoretical 
internal structure constant is a combination of the 
internal structure constants for both star, such that 
\begin{equation}
k_{\rm 2,theo} = \frac{c_{21}k_{21}+c_{22}k_{22}}{c_{21}+c_{22}}
\end{equation}
which can then be compared to observations. 

We find the predicted newtonian apsidal motion to be $\dot \omega_{\rm theo}=$\theowdotnewt~and 
consequently the predicted newtonian internal structure constant to be 
$\log{k_{\rm 2,theo}}=$\theologktwonewt. This is 
in very good agreement with the observed 
$\log{k_{\rm 2,obs}}=$\bestlogktwonewt. 
From equation 9, the parameter $c_{12}$ is about 67\% larger than $c_{22}$.
Therefore, the weighted contribution of
the primary dominates the theoretical apsidal motion. 
V578~Mon is a relatively young system - therefore, $\log{k_{\rm 2,theo}}$ 
is almost constant during the early phases of stellar evolution. 
The ``apsidal motion test'' is therefore complementary to 
the ``isochrone test''. \cite{Claret10} compile a list of eclipsing binaries with 
apsidal motion, demonstrating good agreement between observed and 
predicted apsidal motions. V578~Mon continues this trend of agreement 
between theoretical and observational internal structure constants. For this 
relatively young system, matching the radii, 
temperatures and masses isochrones is key, 
given that we have so few young massive EBs with non-equal mass ratio.

\section{Conclusion}

We have determined the absolute dimensions of the massive,
detached eclipsing binary V578~Mon, which is a member of young 
star forming region NGC 2244 in the Rosette Nebula. We confirm 
that the the previously published spectroscopic orbit 
of \cite{H2000} agree with our current spectroscopic orbit of V578~Mon. 
From our {\sc hermes} spectra, we find that our photometric light ratio from 
the light curve analysis is fully compatible with the disentangled 
component spectra of V578~Mon. 

From 40 yr of Johnson UBV and Str\"{o}mgren $uvby$ photometry we 
determine updated radii, measure the temperature ratio and light ratio 
for the components of V578~Mon. We determine the radii to better 
than $1.5\%$ accuracy, and carefully map out parameter 
space in order to reveal any possible degeneracies. We also 
compare our global best fit light curve model with models that 
include different limb darkening parameters, a different 
assumed temperature for the primary star, light reflection or third light 
finding little effect on our global model. We do not unambiguously rule 
out light reflection or a third body, but we confirm that 
these additional complications to the light curve model will 
not affect our final solution.  

We have compared our observed 
masses, radii, temperatures and rotational velocities to stellar evolution
models of the Geneva, Utrecht, and Granada groups. 
We find no common 
match in predicted ages for mass-radius isochrones of 
the Geneva13 models. We find an age 
difference of only $\gtGenevaDiff$~Myr in predicted ages for the Geneva13 
models for $\log{g}-\log{T_{\rm eff}}$ isochrones.   For the Utrecht11 models, we find a marginally 
common predicted age with an age difference of only \mrUtrechtDiff~Myr for the mass-radius isochrones.  
For the $\log{g}-\log{T_{\rm eff}}$ isochrones we find common ages of $\gtUtrechtDiff$~Myr for the 
Utrecht11 models. For the Granada04 models, 
we find a small age gap of only \mrGranadaDiff~for the mass-radius isochrone, when the primary 
star has a quite large convective overshoot of $\alpha_{\rm ov}=0.6$. We do not find 
common ages for the mass-radius isochrone for the Granada04 models 
when the convective overshoot 
for both stars is a more moderate $\alpha_{\rm ov} = 0.2$. 

This work suggests that models with larger convective overshoot
predict a closer common age
for the components of V578~Mon than models with
a more conventional overshoot of $\alpha_{\rm ov}=0.2$ pressure scale heights.
Evolutionary models with larger convective overshoot extends the size of the convective core for
massive stars, thus extending the main sequence lifetime and allowing for isochrones to
predict a common age for V578~Mon. However - rotational mixing also can prolong
the main sequence lifetime, making the two effects
some what degenerate. The radii may in a small
way be dependent upon
effective temperatures, which are based on imperfect atmosphere
models. Furthermore, there are small systematic residuals of $0.005$
mag in the light curve fits which may in a small way affect the
radii. Finally, effects of binarity, while likely small, are not taken
into account: the side of each star facing the other may be heated and
the addition to the potential $\Omega$ from the
companion is not taken into account into the models. 
The binarity of V578~Mon may cause single 
star models explored here to not be applicable. 

Given the short apsidal period of V578~Mon of 
\uvalue~years, our photometry cover one 
full apsidal motion period. Combined with our
precise measurement of the radii of V578~Mon 
we compute the internal structure constant $\log{k_2}$ 
finding that our observed $\log{k_{\rm 2,obs}}=$\bestlogktwonewt~
in agreement with the theoretical internal structure 
constant $\log{k_{\rm 2,theo}}=$\theologktwonewt. 


V578~Mon is a particularly important system for 
testing stellar evolution models given young age and the difference of 
$\approx30\%$ in the masses of the primary and secondary 
component star. B-type detached 
eclipsing binaries such as V1388 Ori and V1034 Sco 
have similar differences in mass of 40\% and 50\% 
respectively, meaning these systems are also of particular 
importance to providing constraints on 
stellar evolution models. However, V578~Mon is unique 
among such systems by virtue of its young age, thus 
providing the strongest constraints on the models at the 
earliest stages of massive stellar evolution. 

Future work may include comparing the carefully vetted sample 
of high mass EBs in the \cite{Torres10} sample to evolutionary models, include 
more recent massive EBs such as V 380 Cyg \citep{Tka14} and 
LMC 172231 and ST2-28 \citep{Massey12}, to see 
if larger convective overshoot parameters allow for common 
predictions of age. 

\acknowledgments
This work was based on observations obtained with the HERMES spectrograph,  
which is supported by the Fund for Scientific Research of Flanders (FWO), 
Belgium, the Research Council of K.U.Leuven, Belgium, the Fonds National
Recherches Scientific (FNRS), Belgium, the Royal Observatory of Belgium, the
Observatoire de Gen\`{e}ve, Switzerland and the ThŸringer Landessternwarte
Tautenburg, Germany. This work was also conducted in part using the resources of the 
Advanced Computing Center for Research and Education (ACCRE) 
at Vanderbilt University, Nashville, TN. The authors would like 
to acknowledge helpful comments from the referee that improved the paper.

\begin{deluxetable}{ccc}
\tablecolumns{3}
\tablewidth{0pc} 
\tablecaption{\label{Table:obs} 
Identifications, location and combined photometric parameters for eclipsing binary V578~Mon}
\tablehead{} 
\startdata
\colhead{} & \colhead{V578~Mon} & \colhead{Reference} \\
Henry Draper number        &  HD 259135          &   \cite{CP23} \\
Bonner Durchmusterung        & BD\,+04\degr 1299     & \cite{Arge03}    \\ 
Hoag number                & NGC 2244 200           &   \cite{Hog98} \\ \hline 
$\alpha_{2000}$              &   06 32 00.6098        &  \cite{Hog98}  \\
$\delta_{2000}$            & + 04 52 40.902         &   \cite{Hog98} \\
Spectral type             & B0V + B1V & \cite{H2000}   \\ \hline
$V$                       & 8.542                & \cite{OI81}    \\
$V-I$ & 0.262 & \cite{Wang08}  \\
$B-V$                       &+ 0.165               & \cite{OI81}   \\
$U-B$                    &$-$ 0.727                  & \cite{W07}    \\ 
$V-R$                    &+ 0.452               &   \cite{Wang08} \\
\enddata
\end{deluxetable}

\begin{deluxetable}{lllllllll}
\tablecolumns{7}
\tablewidth{0pt} 
\tablecaption{\label{Table:Phot} V578 Mon Light Curves}
\tablehead{ 
\colhead{Observatory} & \colhead{Year} & \colhead{Filter}& \colhead{$\sigma_{0}$} & \colhead{$\sigma$} & \colhead{N} \\
\colhead{} & \colhead{} & \colhead{} & \colhead{[mag]} &\colhead{[mag]} & \colhead{} 
}
\startdata 
$^{1}$ KPNO & 1967-84 & Johnson $U$ &  0.004 & $0.016 $& 251\\ 
& & Johnson $B$ & 0.004 & $0.012 $ & 256 \\ 
& & Johnson $V$ &  0.004 & $0.013 $ & 217 \\ 
$^{2}$SAT & 1991--94& Str\"{o}mgren $u$ & 0.0029 & 0.0067 & 248 \\
& & Str\"{o}mgren $b$ &  0.0023 &  0.0046& 248 \\
& & Str\"{o}mgren $v$ &  0.0023 &  0.0054 & 248 \\
& & Str\"{o}mgren $y$ &  0.0030 &  0.0053 & 248  \\ 
$^{3}$APT & 1994--95 & Johnson $V$ & 0.0037 & $0.0022$ &  260 \\ 
&  & Johnson $B$ &  0.001 &  $0.0040$ &  254 \\ 
 APT & 1995--96 & Johnson $V$ & 0.002 &  $0.0035 $ & 95  \\ 
&  & Johnson $B$ & 0.001 & $0.0037 $ & 96  \\
APT &  1999--2000& Johnson $V$ & 0.002 & $0.0058 $ & 259 \\ 
& & Johnson $B$ & 0.001 & $0.0078 $ & 246  \\
APT & 2005--06 & Johnson $V$ & 0.002 & $0.0036$ & 284  \\ 
& & Johnson $B$ & 0.001 & $0.0044$  & 283 \\
\enddata
\tablecomments{ $^{1}$16-inch telescope at Kitt Peak (KPNO)\\
$^{2}$0.5 m telescope at La Silla (SAT)\\
$^{3}$TSU-Vanderbilt 16-inch telescope at Fairborn University (APT)\\}
\end{deluxetable}

\begin{deluxetable}{ccc}
\tabletypesize{\scriptsize}
\tablecolumns{3}
\tablewidth{0pt} 
\tablecaption{\label{Table:Hermes} {\sc Hermes} Observations}
\tablehead{ 
\colhead{Phase} & \colhead{BJD-2450000.000} & \colhead{Exp Time [s]} 
}
\startdata 
0.9957 & 5904.586 & 2100 \\
0.0060 & 5904.611 & 2100 \\
0.0168 & 5904.637 & 2100 \\
0.0272 & 5904.662 & 2100 \\
0.0376 & 5904.687 & 1980 \\
0.0476 & 5904.711 & 1980 \\
0.0613 & 5909.561 & 1500 \\
0.0692 & 5909.580 & 1500\\
0.1128 & 5914.502 & 2100\\
0.1231 & 5914.527 & 2100\\
0.1530 & 5914.599 & 2100\\
0.1634 & 5914.624 & 2100\\
0.2259 & 5907.549 & 2100\\
0.2363 & 5907.574 & 2100\\
0.2803 & 5912.497 & 2100\\
0.2907 & 5912.522 & 2100\\
0.3434 & 5912.649 & 2100\\
0.3534 & 5912.673 & 2100\\
0.4432 & 5905.664 & 2100\\
0.4449 & 5910.485 & 2300\\
0.4536 & 5905.689 & 2100\\
0.4565 & 5910.513 & 2300\\
0.4673 & 5910.539 & 2100\\
0.4777 & 5910.564 & 2100\\
0.5010 & 5910.620 & 2100\\
0.5113 & 5910.645 & 2100\\
0.6427 & 5908.553 & 2200\\
0.6535 & 5908.579 & 2200\\
0.7187 & 5913.553 & 2100\\
0.7291 & 5913.578 & 2100\\
0.7945 & 5906.510 & 2100\\
0.8049 & 5906.535 & 2100\\
0.9278 & 5911.648 & 2200\\
0.9390 & 5911.675 & 2200\\
\enddata
\tablecomments{Time-series {\sc Hermes} spectroscopy of V578~Mon. Each 
exposure is less than 0.01 of the orbital period for V578 Mon 
of $2.4084822$ days. The time series spectra were obtained to cover 
the out-of-eclipse, primary eclipse and secondary eclipse phases. 
}
\end{deluxetable}

\begin{deluxetable}{ccccc}
\tablecolumns{3}
\tablewidth{0pt} 
\tablecaption{\label{Table:RV} Radial Velocity Solutions}
\tablehead{ \colhead{} & \colhead{$q$} & \colhead{$K_1$} & \colhead{$K_2$} & \colhead{$e$} \\ 
\colhead{} & \colhead{} & \colhead{[km s$^{-1}$]} & \colhead{[km s$^{-1}$]} & \colhead{}
}
\startdata 
\cite{H2000} (LC+Spectroscopy) & 0.7078$\pm0.0002$ & 259.8 & 183.9 & 0.0867 \\  
\cite{H2000} RV only  & 0.705$\pm$0.004 & 259.8$\pm$0.8 & 184.4 & 0.0836$\pm0.0008$ \\ 
HERMES Spectra, $e$ fixed &  0.710 & 259.8 & 184.5 & 0.07755  \\ 
HERMES Spectra, $e$ and $\omega$ fixed & 0.709  & 259.4  & 184.0 & 0.07755 \\ 
\enddata
\end{deluxetable}

\begin{deluxetable}{ccc}
\tablecolumns{3}
\tablewidth{0pt} 
\tablecaption{\label{Table:lf} Light Fraction Comparison}
\tablehead{ 
\colhead{Method} & \colhead{Wavelength $\lambda$} & \colhead{Light Fraction $\frac{l_1}{l_1+l_2}$}  \\ 
\colhead{} & \colhead{[nm]} & \colhead{} 
}
\startdata 
Light Curve Analysis (this work) &  Johnson U, 365 & \JohnsUlf \\ 
  & Johnson B, 445 & \JohnsBlf  \\ 
  & Johnson V, 551& \JohnsVlf \\ 
   & Stromgren u, 365& \Stromulf \\ 
  & Stromgren v, 411& \Stromvlf \\ 
  & Stromgren b, 467&  \Stromblf \\ 
  & Stromgren y, 547&   \Stromylf  \\ 
  
 \cite{H2000} & Stromgren v, 411 &  $0.675\pm0.006$ \\ 
  & Stromgren b, 467&  $0.683\pm0.006$ \\
  & Stromgren y, 547&   $0.692\pm0.006$ \\ 

{\sc HERMES} Spectroscopy & 400-500&  \Hermeslf \\ 
\enddata
\end{deluxetable}

\clearpage

\begin{deluxetable}{ccccc}
\tablecolumns{4}
\tablewidth{0pt} 
\tablecaption{\label{table:grid} Light Curve Parameter Ranges Explored}
\tablehead{ 
\colhead{Parameter} & \colhead{Max} & \colhead{Min}  & \colhead{Coarse Grid Spacing} & \colhead{Fine Grid Spacing}
}
\startdata 
Primary Surface Potential, $\Omega_1$ & 5.36 & 4.80 & \OmegaOneSp & \OmegaOneSpF \\  
Secondary Surface Potential, $\Omega_2$ & 5.26 &  4.40 & \OmegaTwoSp & \OmegaTwoSpF \\ 
Inclination, $i$ [deg] & 73.15 & 70.00 & \InclSp & \InclSpF \\ 
Temperature Ratio, $\frac{T_2}{T_1}$ &  0.875 & 0.843 & \TfracSp & \TfracSpF \\ 
\enddata
\end{deluxetable}

\begin{deluxetable}{ccc}
\tablecolumns{3}
\tablewidth{0pc} 
\tablecaption{\label{Table:lc} 
Light Curve Analysis Results and Comparison
}
\tablehead{\colhead{Light Curve Parameters} & \colhead{This Work \& \cite{Garcia11}} &\colhead{H2000}} 
\startdata
Primary Surface Potential, $\Omega_{1}$ & \bestOone  & $5.02\pm0.05$ \\ 
Secondary Surface Potential, $\Omega_{2}$ & \bestOtwo  & $4.87\pm0.06$ \\ 
Temperature Ratio, $\frac{T_2}{T_1}$ & \bestTrat & $0.88\pm0.02$\\
Inclination, $i$ [deg]  &  \bestIncl & $72.58\pm0.3$ \\ 
Eccentricity, $e$ & \eccvalue & \Hensecc \\ 
Angle of Periastron, $w$ [deg]  & $159.8\pm0.33$ & \Hensomega \\
Ephemeris, HJD$_{0}$ [d] & $2449360.6250$ & $2449360.6250$ \\ 
Total Apsidal Motion, $\dot\omega_{\rm tot}$ [deg~cycle$^{-1}$] & \wdotvalue & \\ \hline 
\\
Light Curve Filters & Str\"{o}mgren $uvby$, Johnson $UBV$ & Str\"{o}mgren $uvby$ \\ 
Total Light Curve Points &  $3489$ & $992$ \\ 
\enddata 
\tablecomments{
The uncertainties on light curve parameters $\Omega_1$, $\Omega_2$, $i$ and 
$\frac{T_2}{T_1}$ are determined from confidence intervals in Figure~\ref{fig:degen}. Light curve 
parameters $e$, $w$ and $\dot\omega_{\rm tot}$ are taken from \cite{Garcia11}.  
This work utilizes photometry that span one full apsidal motion period (U$=$\uvalue yr). 
In contrast to the \cite{H2000} analysis, this work incorporates apsidal motion in the light curve model.  
Finally, the temperature ratio from \cite{H2000} is measured from spectral disentangling.  
}
\end{deluxetable}

\begin{deluxetable}{lcccc}
\tablecolumns{7}
\tablewidth{0pt} 
\tablecaption{\label{Table:ld} Limb Darkening Coefficients }
\tablehead{\colhead{Filter} & \colhead{$x_1$} & \colhead{$x_2$}& \colhead{$y_1$} & \colhead{$y_2$}}
\startdata 
\colhead{Square Root Law (adopted)}\\ \hline
Str\"{o}mgren $u$ &     -0.096&    -0.073&     0.631&     0.606\\
Str\"{o}mgren $b$ &     -0.132&    -0.115&     0.672&     0.659\\
Str\"{o}mgren $v$ &     -0.129&    -0.106&     0.607&     0.581\\
Str\"{o}mgren $y$ &     -0.073&    -0.044&     0.612&     0.581\\
Johnson $U$ &     -0.131&    -0.115&     0.685&     0.675\\
Johnson $B$ &     -0.131&    -0.110&     0.654&     0.638\\
Johnson $V$ &     -0.126&    -0.105&     0.602&     0.578\\
\colhead{Linear Law}\\ \hline
Str\"{o}mgren $u$ &      0.282&     0.291&     0.000&     0.000\\
Str\"{o}mgren $b$ &      0.272&     0.281&     0.000&     0.000\\
Str\"{o}mgren $v$ &      0.235&     0.243&     0.000&     0.000\\
Str\"{o}mgren $y$ &      0.293&     0.304&     0.000&     0.000\\
Johnson $U$ &      0.280&     0.291&     0.000&     0.000\\
Johnson $B$ &      0.262&     0.273&     0.000&     0.000\\
Johnson $V$ &      0.235&     0.242&     0.000&     0.000\\
\colhead{Logarithmic Law}\\ \hline
Str\"{o}mgren $u$ &      0.450&     0.452&     0.252&     0.242\\
Str\"{o}mgren $b$ &      0.450&     0.457&     0.268&     0.264\\
Str\"{o}mgren $v$ &      0.397&     0.398&     0.242&     0.233\\
Str\"{o}mgren $y$ &      0.456&     0.459&     0.244&     0.232\\
Johnson $U$ &      0.462&     0.471&     0.274&     0.270\\
Johnson $B$ &      0.436&     0.444&     0.261&     0.256\\
Johnson $V$ &      0.395&     0.396&     0.241&     0.231\\
\enddata
\tablecomments{Our best fit model uses the square root limb darkening law. Fits with 
the linear cosine or logarithmic limb darkening law had little effect on our final 
light curve solution.}
\end{deluxetable}

\begin{deluxetable}{cccccccc}
\tablecolumns{8}
\tablewidth{0pt} 
\tablecaption{A comparison of light curve models \label{Table:Models}}
\tablehead{ 
\colhead{Model} & \colhead{$\Omega_1$} & \colhead {$\Omega_2$} & \colhead{$i$} & \colhead{$\frac{T_2}{T_1}$} & \colhead{$\chi^2$}  \\ 
\colhead{} & \colhead{} & \colhead{} & \colhead{[deg]} & \colhead{} & \colhead{} 
}
\startdata 
Best Fit &  \bestOone & \bestOtwo &  \bestIncl & \bestTrat & \bestchisqr  \\ 
Fitting for LD coefficients &  \ldOone & \ldOtwo &  \ldIncl & \ldTrat & \ldchisqr  \\ 
Linear Law  & \cosOone & \cosOtwo &  \cosIncl & \cosTrat & \coschisqr  \\ 
Logarithmic Law  & \logOone & \logOtwo &  \logIncl & \logTrat & \logchisqr  \\ 
 Fix $T_1 = 28500$ & \TLowOone & \TLowOtwo &  \TLowIncl & \TLowTrat & \TLowchisqr  \\ 
 Fix $T_1 = 31500$    & \ThiOone & \ThiOtwo &  \ThiIncl & \ThiTrat & \Thichisqr  \\ 
Light Reflection & \refOone & \refOtwo &  \refIncl & \refTrat & \refchisqr  \\ 
Third Light & \ThirdOone & \ThirdOtwo &  \ThirdIncl & \ThirdTrat & \Thirdchisqr  \\ 
\enddata
\tablecomments{The best fit model uses the square root limb darkening law, 
a fixed $T_1=$30000 K, no light reflection, 
and no third light. }
\end{deluxetable}

\begin{deluxetable}{cccc}
\tablecolumns{7}
\tablewidth{0pt} 
\tablecaption{\label{Table:l3} Third Light }
\tablehead{ 
\colhead{Observatory} & \colhead{Year} & \colhead{Filter}& \colhead{$\frac{L_3}{L_{\rm tot}}$}
}
\startdata 
APT  &  2005--06  &  Johnson $B$  &     0.0441 \\
\vphantom  &  \vphantom  &  Johnson $V$  &     0.0218 \\
APT  &  1999--2000  &  Johnson $B$  &     0.0158 \\
\vphantom  &  \vphantom  &  Johnson $V$  &     0.0080 \\
APT  &  1995--96  &  Johnson $B$  &    -0.0037 \\
\vphantom  &  \vphantom  &  Johnson $V$  &     0.0104 \\
APT  &  1994--95  &  Johnson $B$  &     0.0059 \\
\vphantom  &  \vphantom  &  Johnson $V$  &     0.0046 \\
SAT  &  1991--94  &  Str\"{o}mgren $u$  &    -0.0116\\
\vphantom  &  \vphantom  &  Str\"{o}mgren $v$  &    -0.0004 \\
\vphantom  &  \vphantom  &  Str\"{o}mgren $b$  &     0.0013 \\
\vphantom  &  \vphantom  &  Str\"{o}mgren $y$  &    -0.0045 \\
KPNO  &  1967-84  &  Johnson $U$  &     0.0163 \\
\vphantom  &  \vphantom  &  Johnson $B$  &     0.0467 \\
\vphantom  &  \vphantom  &  Johnson $V$  &    -0.0100 \\
\enddata
\tablecomments{Our best fit light curve model includes no third light. 
The small amount of third light varies as a function of epoch.}
\end{deluxetable}


\begin{deluxetable}{lcc}
\tablecolumns{3}
\tablewidth{0pc} 
\tablecaption{\label{Table:absdim} 
V578 Mon Absolute Dimensions
}
\tablehead{
\colhead{Parameter} & \colhead{Primary} & \colhead{Secondary}
}
\startdata
Orbital Period, P [d]  & \multicolumn{2}{c}{$2.4084822$}  \\ 
Mass, $M$ [$M_{\odot}$] & \bestMone & \bestMtwo \\ 
Radius, $R$ [$R_{\odot}$] & \bestRone & \bestRtwo  \\ 
Effective Temperature, $T_{\rm eff}$ [K] & \bestTone & \bestTtwo  \\
Surface Gravity, $\log{g}$ [cm s$^{-2}$] & \bestloggone  & \bestloggtwo  \\ 
Surface Velocity, $v_{\rm rot}$ [km s$^{-1}$] & \bestvone & \bestvtwo \\
Luminosity, $\log{\frac{L}{L_{\odot}}}$ & \bestLone & \bestLtwo \\
Synchronicity Parameter, $F=\frac{w}{w_{orb}}$ & \bestFone & \bestFtwo \\ 
Apsidal Period, U [yr] &  \multicolumn{2}{c}{\uvalue} \\ 
Observed Newtonian Internal Structure Constant, $\log{k_{\rm 2,newt}}$ & \multicolumn{2}{c}{\bestlogktwonewt} \\
\enddata
\end{deluxetable}

\clearpage
\begin{deluxetable}{cccc}
\tablecolumns{4}
\tablewidth{0pt} 
\tablecaption{\label{Table:evol} Stellar Evolution Models Comparison
}
\tablehead{ 
\colhead{Physical Input} & \colhead{Geneva13} & \colhead{Utrecht11} & \colhead{Granada04} 
}
\startdata


Composition  [Z,Y,X] &  [0.014,0.266,0.720]  &  [0.0122,0.2486,0.7392]  & [0.014,0.271,0.715] \\
Overshoot, $\alpha_{\rm ov}$ & 0.10   & 0.355 & 0.6 pri, 0.2 sec \\  
Mixing Length, $\alpha_{\rm MLT}$ & 1.60 & 1.5 & 1.68 \\  
Rotation & Yes & Yes & Yes \\
Rotational Mixing & Yes  & Yes & No    \\ 
Opacities & \cite{IglesiasRogers96} & \cite{IglesiasRogers96} & \cite{IglesiasRogers96} \\ 
Mass loss  & \cite{Vink01} & \cite{Vink01} & \cite{Vink01} \\

\enddata
\end{deluxetable}

\clearpage
\begin{deluxetable}{lcccc}
\tablecolumns{3}
\tablewidth{0pt} 
\tablecaption{\label{Table:ages} Ages from Stellar Evolution Models}
\tablehead{ 
\colhead{Model} & \colhead{Primary Age} & \colhead{Secondary Age} & \colhead{Age Diff (lower limit)} & \colhead{$\alpha_{\rm ov}$}\\
\colhead{} & \colhead{[Myr]} & \colhead{[Myr]} & \colhead{[Myr]} & \colhead{[Scale height]} 
} 
\startdata



\multicolumn{5}{c}{Mass$-$Radius$-v_{\rm rot}$ Isochrones} \\
Geneva13 & 4.3$-$4.6&6.2$-$7.1 & \mrGenevaDiff  & 0.1 \\
Utrecht11 &3.0$-$3.2&3.6$-$4.4 & \mrUtrechtDiff  & 0.355\\
Granada04 & 5.0$-$5.3&5.5$-$6.3 &  \mrGranadaDiff & 0.6 pri, 0.2 sec \\

\multicolumn{5}{c}{$\log{g}$-$\log{T_{\rm eff}}$-$v_{\rm rot}$ Isochrones} \\
Geneva13 & 3.9$-$5.1&5.2$-$7.5 & \gtGenevaDiff & 0.1 \\
Utrecht11  & 2.6$-$3.8&2.4$-$5.2 & Common age $3.5\pm1.5$ & 0.355 \\
Granada04 & 4.7$-$5.5&4.9$-$6.8 & Common age $5.5\pm1.0$ &  0.6 pri, 0.2 sec  \\
\enddata

\tablecomments{The ages for the primary and secondary star are computed from evolutionary 
tracks at the masses of either star and solar metallicity. The Granada04 models were 
computed for a high convective overshoot of $\alpha_{\rm ov}=0.6$ pressure scale heights for the primary star, which allowed 
the models to match the observations. It easier to find a common age for the $\log{g}-\log{T_{\rm eff}}$ 
isochrone given the larger uncertainty on the effective temperatures of the stars. 
}
\end{deluxetable}

\end{document}